\documentclass[aps,pra,reprint,onecolumn,11pt,superscriptaddress,longbibliography]{revtex4-1}
\usepackage[colorlinks=true,allcolors=blue]{hyperref}
\usepackage{amsmath}
\usepackage{amssymb}
\usepackage{bm}
\usepackage[pdftex]{graphicx}
\usepackage{microtype}
\usepackage{verbatim}
\usepackage{todonotes}
\usepackage{caption}
\usepackage{subcaption}
\usepackage{makecell,tabularx}
\presetkeys{todonotes}{inline,backgroundcolor=yellow}{}
\usepackage{subcaption}
\usepackage{soul}
\newcommand{\bp}[1]{\left( {#1} \right)} % "bracket-parenthetical"
\newcommand{\bs}[1]{\left[ {#1} \right]} % "bracket-square"
\newcommand{\ket}[1]{{| {#1} \rangle}}

\newcommand{\ketbra}[2]{{\left| {#1} \right\rangle \!\!\left\langle {#2} \right|}}

\newcommand{\cg}[6]{
\left\langle 
\begin{array}{@{}cc|c@{}}
    #1 & #3 & #5 \\
    #2 & #4 & #6
\end{array}
\right\rangle
}

\DeclareMathOperator{\tr}{tr}
\DeclareMathOperator{\Sym}{Sym}

\begin{document}
%\linenumbers
\title{Stellar representation of extremal Wigner-negative spin states}
\author{Jack Davis}
\affiliation{Institute for Quantum Computing, University of Waterloo, Waterloo, Ontario, Canada N2L 3G1}
\affiliation{Department of Physics and Astronomy, University of Waterloo, Waterloo, Ontario, Canada N2L 3G1}
\author{Robie A. Hennigar}
\affiliation{Departament de F{\'\i}sica Qu\`antica i Astrof\'{\i}sica, 
 Universitat de
Barcelona (UB), c.~Mart\'{\i} i Franqu\`es 1, E-08028 Barcelona, Spain
}
\affiliation{Institut de Institut de
Ci\`encies del Cosmos, (ICCUB), Universitat de Barcelona (UB), c.~Mart\'{\i} i Franqu\`es, 1, E-08028 Barcelona, Spain.}
\author{Robert B. Mann}
\affiliation{Department of Physics and Astronomy, University of Waterloo, Waterloo, Ontario, Canada N2L 3G1}
\affiliation{Institute for Quantum Computing, University of Waterloo, Waterloo, Ontario, Canada N2L 3G1}
\affiliation{Perimeter Institute for Theoretical Physics, 31 Caroline St N, Waterloo, Ontario, Canada N2L 2Y5}
\author{Shohini Ghose}
\affiliation{Department of Physics and Computer Science, Wilfrid Laurier University, Waterloo, Ontario, Canada N2L 3C5}
\affiliation{Institute for Quantum Computing, University of Waterloo, Waterloo, Ontario, Canada N2L 3G1}
\affiliation{Perimeter Institute for Theoretical Physics, 31 Caroline St N, Waterloo, Ontario, Canada N2L 2Y5}

\begin{abstract}
The Majorana stellar representation is used to characterize spin states that have a maximally negative Wigner quasiprobability distribution on a spherical phase space.  These maximally Wigner-negative spin states generally exhibit a partial but not high degree of symmetry within their star configurations.  In particular, for spin $j > 2$, maximal constellations do not correspond to a Platonic solid when available and do not follow an obvious geometric pattern as dimension increases.  In addition, they are generally different from spin states that maximize other measures of nonclassicality such as anticoherence or geometric entanglement.  Random states ($j \leq 6$) display on average a relatively high amount of negativity, but the extremal states and those with similar negativity are statistically rare in Hilbert space. We also prove that all spin coherent states of arbitrary dimension have non-zero Wigner negativity.  This offers evidence that all pure spin states also have non-zero Wigner negativity.  The results can be applied to qubit ensembles exhibiting permutation invariance.
\end{abstract}

\maketitle
\tableofcontents

\section{Introduction}

The rise of quantum-enhanced technologies for information processing, communication, metrology, and machine learning has driven   efforts to better characterize the quantum resources needed to power them.  Highly nonclassical states are critical for delivering improved performance in these applications.  For example, states with a high degree of entanglement are crucial for tasks such as teleportation, entanglement swapping, and quantum key distribution \cite{Vidal_2003}.  However the presence of a large amount of entanglement is not necessarily a universal source of ``quantumness" that powers the applications of quantum theory \cite{universal_comp_no_ent_vandenNest_2013}; alternative definitions of nonclassicality have been proposed \cite{Brodutch_Terno_discord_2017, Howard_Wallman_Veitch_Emerson_2014}.  To have a complete picture of the nonclassicality of quantum resources---and hence a better understanding of states likely to supply a quantum advantage---it is worthwhile to categorize inequivalent notions of nonclassicality and determine which states extremize them.  Wigner negativity, defined as the negative volume of the Wigner distribution representing a quantum state on phase space, is such an alternative measure of nonclassicality \cite{Kenfack_2004}.  Here we investigate quantum spin states that minimize or maximize SU(2)-covariant Wigner negativity.

Previous studies of negative quasiprobability in the continuous Heisenberg-Weyl Wigner function and its generalizations have been fruitful.  Pure states in the canonical phase space representation with Gaussian field quadratures (equivalently those with positive Wigner functions \cite{Hudson_1974,soto1983wigner}) are known to be efficiently simulatable on a classical computer, an extension of the Gottesman-Knill theorem to the continuous variable regime \cite{Bartlett_Sanders_2002,Mari_Eisert_2012,Albarelli_2018}.  In this sense, negative quasiprobability is understood as a necessary resource for continuous variable quantum computing, though other protocols exist where non-Gaussianity also leads to improved performance \cite{Albarelli_2018}.  The use of phase space methods to characterize quantum resources analogously exists in the discrete variable regime, particularly with the magic state injection model of quantum computation.  The magic state scheme is one wherein stabilizer operations alone may distill certain non-stabilizer states to a resourceful ``magic" state, capable of elevating the stabilizer framework to universality \cite{Bravyi_Kitaev_2005}.  Exactly which non-stabilizer states are capable of doing so were identified by Veitch \textit{et al}.\ using finite-dimensional phase space methods \cite{Veitch_2012}.  In particular, states with Wigner negativity in the Gross-Wigner phase space picture were necessary for magic distillation \cite{Veitch_resource_2014}.

Relatively less research has been done on the negativity of other generalized Wigner functions, particularly those not associated with a Heisenberg-like dynamical symmetry.  Such non-Heisenberg Wigner functions are well-defined through the parity-operator framework, known also as the Stratonovich framework \cite{stratonovich_distributions_1956, Varilly_Bondia_1989, Brif_Mann_1999}.  In this phase space picture the dynamical symmetry group $G$ plays a central role, with all quasiprobability distributions required to be $G$-covariant.  The canonical Wigner function and the Gross-Wigner function are special cases of the Stratonovich-Wigner function, with $G$ respectively being the continuous Heisenberg-Weyl group \cite{royer_wigner_1977} or the Pauli group (equivalently the discrete Heisenberg-Weyl group) \cite{Gross_2006, Klimov_2006}.  Little work has been done on the negativity of other Stratonovich-Wigner functions, particularly so in the simplest non-Heisenberg case of $G=$ SU(2) \cite{Klimov_Romero_deGuise_2017, Heiss_Weigert_discrete_2000, Koczor_continuous_2020}.  This dynamical symmetry corresponds to a spherical phase space, and the associated Wigner function is well-defined in all finite dimensions, including a natural description of the single qubit.  This is in contrast to the Gross-Wigner function, which is only defined in odd dimensions.

Physically realizable systems that exhibit SU(2) symmetry include trapped ions \cite{Leibfried_trapped_6q_cat_2005}, large atomic ensembles \cite{McConnell_3000_atoms_2015}, orbital angular momentum photon states \cite{Bouchard_2017}, photon polarization states \cite{Rozema_2014_biphoton}, and any system mathematically equivalent to the symmetric subspace of a collection of qubits \cite{harrow2013church}.  Spin systems are additionally amenable to the Majorana stellar representation, which provides a faithful visualization of spin states with arbitrarily high dimension \cite{Majorana_1932}.  In particular, pure states with spin $j$ are uniquely represented as a constellation of $2j$ points on the unit sphere.  Using this representation, the general task of characterizing the most nonclassical spin state is equivalent to the task of characterizing a spherical distribution of points -- a historic mathematical problem with many acceptable solutions \cite{Whyte_1952, saff1997distributing}.  Several notions that quantify the nonclassicality of angular momentum have been studied in this stellar context, including anticoherence \cite{zimba2006anticoherent,bjork_kings_2015}, $P$-representability \cite{Giraud_2008_classicality, Giraud_2010}, and the geometric measure of entanglement \cite{martin_giraud_geometric_high_2010, Aulbach_2010}; see also \cite{mixed-1qubit-reductions_bastin_2014, Goldberg_2020_extremal, Goldberg_Grassl_Leuchs_Sanchez-Soto_2021}.

In this article we analyze constellations that are highly nonclassical with respect to the SU(2)-covariant Stratonovich-Wigner function.  In the cases of $j \leq \frac{7}{2}$ we numerically calculate the constellation with the globally maximal Wigner negativity.  These constellations are unique up to rigid rotation.  For spins $4 \leq j \leq 6$ we switch to a statistical analysis based on randomly sampling Hilbert space according to the circular unitary ensemble. In general, the constellations of the maximally/highly Wigner-negative states do not exhibit standard point group symmetries. We compare these constellations to those that maximize other measures of nonclassicality.  We find partial agreement with these alternative maximal constellations in spin systems going up to $j=2$, followed by complete disagreement for $j > 2$.  In the high-spin cases where the Wigner-maximal state was not determined, the most negative randomly sampled state is nonetheless found to be more negative than extremal states corresponding to the other nonclassicality measures.  This is especially notable in the cases of $j=3$ and $j=6$ where the alternative measures collectively identify the so-called ``octahedron" and ``icosahedron" states as extremal, whereas in both cases we find a different state with higher negativity.  Unlike the Platonic states, the states we identify -- exactly in the former and statistically in the latter -- are relatively irregular in their structure and do not adhere to an obvious geometric principle.  We argue that a simple analysis based on either constellation symmetry or constellation delocalization is insufficient to characterize how Wigner negativity scales with dimension.  We also generally find that the average state across Hilbert space has a relatively large amount of negativity, but that it is rare to have a negativity close to the maximal value.

In an independent result, we also present a proof that all spin coherent states of arbitrary dimension must have non-vanishing SU(2)-covariant negativity.  This suggests that all pure spin states also have non-vanishing SU(2)-covariant negativity. If true, this property would be unique to the spherical phase space as compared to the planar or toroidal lattice phase spaces of Heisenberg-Weyl symmetry.  Statistical analysis offers further evidence towards this claim.

We begin with a review of the stellar representation and the Stratonovich framework.  We then discuss our results, beginning with the proof that all spin coherent states of arbitrary dimension have a spherical Wigner function that must take negative values.  This is followed by an analysis of the maximally Wigner-negative spin states for $j$ ranging up to $\frac{7}{2}$ (equivalent to seven indistinguishable qubits).  This includes notable non-extremal cases, and a more in-depth treatment for the cases of $j=\frac{1}{2},1,\frac{3}{2}$ on account of the low dimensionality.  A statistical analysis is then done on uniformly sampled random states.  Comparisons are made throughout to the following three alternative definitions of nonclassicality discussed in the context of spin: anticoherence (i.e.\ Kings of Quantum) \cite{zimba2006anticoherent,bjork_kings_2015}, $P$-representability (i.e.\ Queens of Quantum) \cite{Giraud_2008_classicality, Giraud_2010}, and the geometric measure of entanglement (non-regal) \cite{martin_giraud_geometric_high_2010, Aulbach_2010}.  We end with a general discussion and conclusion.  Appendix \ref{sec:alternative_descriptions} summarizes the alternative measures of nonclassicality and Appendix \ref{sec:numerical_data} contains numerical data of the maximal Wigner-negative states.
 
\section{Background}\label{sec:background}

\subsection{Stellar representation}
The stellar representation of spin, attributed to Majorana, is an extension of the Bloch sphere, offering a description of a pure spin-$j$ state as a unique constellation of $2j$ points on the unit sphere \cite{Majorana_1932}.  Several applications have been established, including a generalization of the geometric phase \cite{Hannay_1998, Bruno_geometric_2012} and the classification of symmetric state entanglement with respect to stochastic local operations and classical communication (SLOCC) \cite{Bastin_operational_families_2009,Bastin_symmetric_ILO_2010, Markham_nonlocality_hardy_2012}.  We briefly review the stellar representation without specific derivations.  See Refs.\ \cite{Chryssomalakos_geometry_2018, Devi_2012} for more information.

The physical state space of a two level quantum system is the projective Hilbert space under $\mathbb{C}^2$, topologically understood as the extended complex plane $\mathcal{P}\mathbb{C}^1 = \mathbb{C} \cup \{\infty\}$.  This is isomorphic to the Riemann sphere $S^2$, with an explicit bijection given by the inverse stereographic projection to a sphere centred at the origin.  The projection point is by convention the South pole, which is where the point at infinity is mapped.  Given a qubit state in the standard polar coordinate parameterization $\ket{\psi} = \cos\frac{\theta}{2}\ket{0} + e^{i\phi}\sin\frac{\theta}{2}\ket{1}$, the ratio between the spin-down and spin-up amplitudes, $\tan\frac{\theta}{2}e^{i\phi}$, is the associated point in $\mathcal{P}\mathbb{C}^1$.  This is the famous Bloch sphere picture of a qubit.  The key insight to the stellar perspective is that a state within a higher dimensional irrep of SU(2) may be characterized by an \textit{unordered set} of points in $\mathcal{P}\mathbb{C}^1$ rather than a single point in a larger space.  Thus the correspondence between $\mathcal{P}\mathbb{C}^1$ and $S^2$ allows the identification of spin states with constellations on the sphere; simply perform the inverse stereographic projection on each point in $\mathcal{P}\mathbb{C}^1$ comprising the given state.  The number of stars in the constellation is twice that of the spin, $n=2j$, counting multiplicities.

On the other hand, any spin-$j$ state must be expressible as a linear combination of the angular momentum (Dicke) basis,
\begin{equation}
    \ket{\Psi} = \sum_{m=-j}^j a_m \ket{j,m}.
\end{equation}
The connection between the algebraic and geometric descriptions is supplied by the zero-set of the Majorana polynomial, 
\begin{equation}\label{eq:Majorana_poly}
    P_{\ket{\Psi}}(z) = \sum_{m=-j}^j (-1)^{j-m} \sqrt{\binom{2j}{j-m}} a_m z^{j+m}.
\end{equation}
This is the polynomial over $\mathbb{C}$ with roots given by the non-infinite points in $\mathcal{P}\mathbb{C}^1$ characterizing a quantum state,
\begin{equation}
    P_{\ket{\Psi}}(z) = a_j \prod_{i=1}^{2j} (z - z_i) \qquad z_i = \tan\frac{\theta_i}{2}e^{i\phi_i} \neq \infty
\end{equation}
with $a_j = \langle j,j | \Psi \rangle$ being the leading coefficient.  If the number of roots is less than $2j$ then the remaining roots are ``at infinity", and are associated with the South pole (i.e.\ $|z_i| \rightarrow \infty \Rightarrow \theta_i \rightarrow \pi$).

The stellar representation has a natural interpretation when viewing a spin-$j$ system as the symmetric subspace of the Hilbert space of $n=2j$ qubits, $\Sym^n (\mathbb{C}^2) \simeq \mathbb{C}^{n+1} \subset \mathbb{C}^{2^n}$.  Consider the tensor product of a set of $n$ qubits
\begin{equation}\label{eq:general_qubit_product}
\ket{\Psi} = \bigotimes_{i=1}^n \ket{\psi_i}, \qquad \ket{\psi_i} = \cos\frac{\theta_i}{2} \ket{0} + e^{i\phi_i}\sin\frac{\theta_i}{2}\ket{1}
\end{equation}
where $(\theta_i, \phi_i)$ are the respective Bloch vectors.  This state lives in the Hilbert space $(\mathbb{C}^2)^{\otimes n}$ and is in general not invariant under permutations of the individual qubits; they are distinguishable.  Permutation invariance (indistinguishability) is enforced by application of the symmetrizer, $\pi_n: (\mathbb{C}^2)^{\otimes n} \rightarrow (\mathbb{C}^2)^{\otimes n}$,
\begin{equation}\label{eq:symmetrizer}
    \pi_n := \frac{1}{n!} \sum_{\tau \in S_n} R_\tau,
\end{equation}
where $R_\tau$ is the irreducible unitary representation of the symmetric group of degree $n$, and acts on the qubit indices of $\ket{\Psi}$.  The symmetrizer $\pi_n$ is the orthogonal projection operator onto the symmetric subspace of $n$ qubits, which furthermore forms an irreducible representation of SU(2) with spin $j=n/2$ \cite{harrow2013church}.  The resulting spin state $\ket{\Psi_S} \propto \pi_n \ket{\Psi}$ is the state associated with the constellation defined by the Bloch vectors $\{ (\theta_i, \phi_i) \}$ in Eq.\ \eqref{eq:general_qubit_product}, whose stereographic projections form the zero set of the Majorana polynomial $P_{\ket{\Psi_S}} (z)$.  A useful property of the qubit picture is that a rigid rotation of a given constellation, $e^{i \theta \boldsymbol{n} \cdot \boldsymbol{J}}$, amounts to the simultaneous local rotation of each constituent qubit by $e^{i \theta \boldsymbol{n}\cdot \frac{\boldsymbol{\sigma}}{2}}$.

For example, consider the Dicke constellations.  Begin with the state of a distinguishable ensemble of $2j$ qubits with the last $k$ spin-down along $z$, 
\begin{equation}\label{eq:Dicke_const_example}
    \ket{\psi} = | \underbrace{0\cdots 0}_{2j-k} \,  \underbrace{1\cdots 1}_{k} \rangle.
\end{equation}
Symmetrize and renormalize:
\begin{equation}
\begin{aligned}
    \pi_{2j} \ket{\psi} &= \frac{1}{(2j)!} \sum_{\tau \in S_{2j}} | \tau (\underbrace{0\cdots 0}_{2j-k} \,  \underbrace{1\cdots 1}_{k} ) \rangle \\
    &\mapsto \binom{2j}{k}^{-\frac{1}{2}} \frac{1}{(2j-k)!k!} \sum_{\tau \in S_{2j}} | \tau (\underbrace{0\cdots 0}_{2j-k} \,  \underbrace{1\cdots 1}_{k} ) \rangle \\
    &:= \ket{j,j-k}.
\end{aligned}
\end{equation}
The Majorana polynomial of this state is
\begin{equation}
    P_\ket{j,j-k}(z) = \sum_{m=-j}^j (-1)^{j-m} \sqrt{\binom{2j}{j-m}} \delta_{m, j-k} z^{j+m} = (-1)^k \sqrt{\binom{2j}{k}} z^{2j-k}.
\end{equation}
This is a monomial with a $(2j-k)$-degenerate zero at $z=0$. Thus there are $(2j-k)$ stars on the North pole via the stereographic map.  The remaining $k$ stars must then be on the South pole, matching the original Bloch vectors in Eq.\ \eqref{eq:Dicke_const_example}.

\subsection{Stratonovich phase space representation of spin}
The quantum phase space picture is usually restricted to systems characterized by continuous Heisenberg-Weyl symmetry $H_d$ in $d$ spatial dimensions, equivalently $d$ bosonic modes.  The Weyl rule maps operators to functions over the phase space $\mathbb{C}^d$, and different operator $s$-orderings yield different functional representations \cite{Cahill_Glauber_1, Cahill_Glauber_2}.  A natural generalization of this picture is attributed to Stratonovich, who postulated a set of axioms applicable to systems characterized by SU(2) dynamical symmetry \cite{stratonovich_distributions_1956, Varilly_Bondia_1989}.  It has since been generalized to a larger class of symmetries, where the phase space $\Gamma_G$ is derived from the group manifold $G$, typically as a quotient space \cite{Brif_Mann_1999}.  The generalized $s$-ordered quasiprobability distribution of any operator $A$ is defined through an operator-valued distribution $\Delta^{(s)}_{G}: \Gamma_G \rightarrow \mathcal{L}(\mathcal{H})$ on phase space:
\begin{equation}\label{eq:weyl_rule}
    f^{(s)}_{A} (\Omega) := \tr[A \Delta^{(s)}_{G}(\Omega)], \qquad \quad \Omega \in \Gamma_G.
\end{equation}
Eq.\ \eqref{eq:weyl_rule} is the generalized Weyl rule and the values of $\Delta^{(s)}_{G}$ are called phase-point operators or the $s$-ordered $G$-kernel.  For example, the single-mode optical Wigner function ($G=H_1$, $s=0$) is given by
\begin{equation}
W_\rho (\alpha) := f^{(0)}_\rho = \tr[\rho \,\Delta^{(0)}_{H_1}(\alpha)] \, \, , \qquad \qquad \Delta^{(0)}_{H_1} (\alpha) = 2 D(\alpha) e^{i\pi N} D^\dagger (\alpha)
\end{equation}
where $\Gamma_{H_1} = \mathbb{C} \simeq H_1/$U(1) is the Heisenberg-Weyl phase space, $\alpha \in \mathbb{C}$, $N=a^\dagger a$, and $D=e^{\alpha a^\dagger - \bar{\alpha}a}$ is the symmetrically ordered displacement operator.  This is equivalent to the common definition of the Wigner function as the Fourier transform of a characteristic function \cite{royer_wigner_1977}.

The phase-point operators are required to satisfy the Stratonovich axioms \cite{Brif_Mann_1999}.  These axioms may be expressed either as properties required by the kernel or by the phase space functions.  On the functional level they are the following:
% \begin{subequations}
% \begin{equation}
%     f^{(s)}_{A^\dagger} (\Omega) = [f^{(s)}_A (\Omega)]^* \qquad \text{realness}
% \end{equation}
% \begin{equation}
%     \tr[A] = \int_{\Gamma_G} d\mu f^{(s)}_A (\Omega) \qquad \text{standardization}
% \end{equation}
% \begin{equation}
%     \tr[AB] = \int_{\Gamma_G} d\mu f^{(s)}_A (\Omega) f^{(-s)}_B (\Omega) \qquad \text{traciality}
% \end{equation}
% \begin{equation}
%     f^{(s)}_{\pi(g) \rho \pi^\dagger (g)}(\Omega) = f^{(s)}_{\rho}(g^{-1}\, \Omega) \qquad \text{covariance}
% \end{equation}
% \end{subequations}
\begin{subequations}
\begin{align}
\begin{split}
    f^{(s)}_{A^\dagger} (\Omega) &= [f^{(s)}_A (\Omega)]^*
\end{split} \quad
\begin{split}
    \text{realness}
\end{split} \\
\begin{split}
     \tr[A] &= \int_{\Gamma_G} d\mu f^{(s)}_A (\Omega)
\end{split} \quad
\begin{split}
    &\text{standardization}
\end{split}\\
\begin{split}\label{eq:traciality}
     \tr[AB] &= \int_{\Gamma_G} d\mu f^{(s)}_A (\Omega) f^{(-s)}_B (\Omega)
\end{split} \quad
\begin{split}
    &\text{traciality}
\end{split}\\
\begin{split}
    f^{(s)}_{\pi(g) \rho \pi^\dagger (g)}(\Omega) &= f^{(s)}_{\rho}(g^{-1}\, \Omega)
\end{split} \quad
\begin{split}
    &\text{covariance}
\end{split}
\end{align}
\end{subequations}
where $\mu$ is the Haar measure on $\Gamma_G$, $g\Omega$ denotes the action of $G$ on $\Gamma_G$, and $\pi$ is an irreducible unitary representation of $G$ acting on $\mathcal{H}$.  Realness and standardization ensure density matrices are mapped to real-valued functions that integrate to unity, and traciality links measurement statistics to phase space averages.  In particular when $s=0$ the traciality and realness axioms, as applied to a state and an observable, equate the Born rule with the $L^2$-inner product of their Wigner functions.  This will come up in the discussion of spin coherent states in Sec.\ \ref{sec:spin_coherent}.  The covariance axiom establishes compatibility between the two group actions; i.e.\ any $G$-action on Hilbert space will have a corresponding $G$-action on phase space (and vice-versa).  The \textit{Wigner negativity} $\delta$ of a state $\rho$ is defined in this general context as:
\begin{equation}
    \label{wigNeg}
    \delta(\rho) = \frac{1}{2}\left( \int_{\Gamma_G} d\mu |f^{(0)}_\rho (\Omega)| -1\right)  .
\end{equation}

We now focus on spin symmetry and restrict attention to pure states.  The phase space associated with $G =$ SU(2) is the sphere with invariant measure $d\mu = \frac{2j+1}{4\pi}\sin\theta d \theta d \phi$ and radius indexed by $j$ \cite{Brif_Mann_1999, Klimov_Romero_deGuise_2017, Koczor_continuous_2020}.  In polar coordinates the spin-$j$ Wigner kernel, denoted from here on as $\Delta_j$, is
\begin{equation}\label{eq:SU(2)-kernel-diagonal}
    \Delta_j (\theta,\phi) = \sum_{m=-j}^j \Delta_{j,m} \ketbra{j,m;\bm{n}}{j,m;\bm{n}}, \qquad \Delta_{j,m} = \sum_{l=0}^{2j} \frac{2l+1}{2j+1} \cg{j}{m}{l}{0}{j}{m}
\end{equation}
where $\bm{n}=(\sin\theta\cos\phi,\sin\theta\sin\phi,\cos\theta)$ is the Stern-Gerlach axis pointing to $(\theta,\phi)$, and $\cg{j_1}{m_1}{j_2}{m_2}{J}{M}$ denotes the Clebsch-Gordon coefficients \cite{Heiss_Weigert_discrete_2000}.  This kernel \eqref{eq:SU(2)-kernel-diagonal} yields the \textit{unique} spherical Wigner function that approaches the CV Wigner function in the limit of infinite spin (i.e.\ radius) \cite{Weigert_Amiet_2000, Koczor_continuous_2020, Arecchi_1972}.  With a fixed quantization about the $z$ axis the kernel is expressed as
\begin{equation}\label{eq:kernel_along_z}
    \Delta_j(\theta,\phi) = \sqrt{\frac{4\pi}{2j+1}} \sum_{K=0}^{2j} \sum_{q=-K}^l T^{(j)}_{Kq} Y^*_{Kq}(\theta,\phi),
\end{equation}
where $Y^*_{Kq}(\Omega)$ is the complex conjugate of the ($\ell=K, m=q$)-spherical harmonic and
\begin{equation}\label{eq:spherical_tensor}
    T^{(j)}_{Kq} = \sqrt{\frac{2K+1}{2j+1}} \sum_{m,m'=-j}^j \cg{j}{m}{K}{q}{j}{m'} \ketbra{j,m'}{j,m}
\end{equation}
are the spherical multipole operators associated with the spin-$j$ unitary irrep on pure states \cite{Klimov_Romero_deGuise_2017}.  For example, again consider the Dicke states along $z$.  Inserting $\rho = \ketbra{j,m}{j,m}$ and Eq.\ \eqref{eq:kernel_along_z} into Eq.\ \eqref{eq:weyl_rule} yields
\begin{equation}\label{eq:dicke-wigner}
    W_{\ket{j,m}}(\theta,\phi) = \sum_{l=0}^{2j} \frac{2l + 1}{2j + 1} \cg{j}{m}{l}{0}{j}{m}
    P_l\bp{\cos\theta}
\end{equation}
where $P_l(x)$ are the Legendre polynomials.

\section{Wigner negativity of spin coherent states} \label{sec:spin_coherent}

A natural question to ask is whether the spin coherent state attains negative values in its Wigner function somewhere on the spherical phase space.  Here we give a simple demonstration of why the spin coherent state Wigner function is not positive semidefinite for arbitrary spin and centroid.  

Hudson's theorem on the planar Wigner function establishes the equivalence between Gaussianity and non-negativity for pure states \cite{Hudson_1974, soto1983wigner}.  As a result, any two non-negative states $\ket{\psi_1}$ and $\ket{\psi_2}$ have non-vanishing overlap, $\langle{ \psi_1 | \psi_2 \rangle} \neq 0$.  
This can be seen from the traciality axiom, Eq.\ \eqref{eq:traciality}, where the associated Wigner functions have non-vanishing tails that will always overlap by some non-zero amount.  The coherent state basis $\{\ket{\alpha}\}$ is a special instance with $\langle{\alpha | \beta\rangle} \neq 0$ for any $\alpha, \beta \in \mathbb{C}$.  For spin systems however there is no analog of Hudson's theorem to yield a similar conclusion; not all pairs of spin coherent states have non-vanishing overlap.  Indeed, any two spin-$j$ coherent states with antipodal centroids are orthogonal.  This follows from the orthogonality of the Dicke basis along any quantization axis $\boldsymbol{n}$, in particular the states with highest and lowest polarization: $\langle{ j, j; \boldsymbol{n} | j, -j; \boldsymbol{n} \rangle} = 0 \, \, \forall \boldsymbol{n} \in S^2$.  With the fully spin-up Dicke state identified with the spin-$j$ coherent state $\ket{\Omega}$ and the fully spin-down with $\ket{\Omega^\perp}$, where $\Omega^\perp = (\theta,\phi)^\perp = (\pi - \theta, \pi + \phi)$ is the coordinate antipodal to $\Omega$, traciality becomes
\begin{equation}\label{eq:antipodal-traciality}
    \frac{2j+1}{4\pi} \int_{S^2} W_{\ket{\Omega}}(\theta,\phi) W_{\ket{\Omega^\perp}}(\theta, \phi) \sin\theta d \theta d \phi = 0.
\end{equation}

The problem is to conclude from Eq.\ \eqref{eq:antipodal-traciality} that all spin coherent states take negative values somewhere in phase space.  This conclusion is not immediate, since Eq.\ \eqref{eq:antipodal-traciality} could \textit{a priori} be satisfied by positive-definite functions with disjoint support.  However this can be ruled out  by appealing to the general form of the Dicke state Wigner function, Eq.\ \eqref{eq:dicke-wigner}.  By restricting to an arc of constant longitude between the two poles, their azimuthal symmetry allows them to be effectively viewed as real-valued functions over the interval $[0,\pi]$.  Furthermore, as they are each a linear combination of Legendre polynomials, which in turn are combinations of powers of their arguments, the Dicke state Wigner function can be seen as a finite-degree polynomial in $\cos\theta$.  Being such a polynomial with real coefficients, their real zero sets will be finite, hypothetically ranging from no roots to $2j$ roots \footnote{It was pointed out in Ref.\ \cite{Davis_Wigner_2021} that there is no clear relationship between the spin $j$ and the number of distinct roots in the Dicke state Wigner functions}.  So any two such polynomials are supported on $[0,\pi]$ up to some finite set of points (i.e.\ measure zero), implying that they will always overlap for finite $j$.  Traciality \eqref{eq:antipodal-traciality} then forces at least one of the two Wigner functions to take negative values somewhere.  This implies that either the North pole or South pole spin coherent state is negative somewhere.  But since any two spin coherent states are connected through a rigid rotation, which preserves negativity, all spin coherent states must then also take negative values.

The above result, together with the conjecture that spin coherent states also minimize SU(2)-covariant Wigner negativity, would imply that all pure spin states are nonclassical in some sense.  This likely non-existence of Gaussian/stabilizer analogues may then reflect the nonclassicality of spin symmetry itself, and so offer some intuition for why quantum mechanical spin has no classical analogue.

\section{Small spins}
Here we discuss spins systems with $j=1/2, 1, 3/2$, and the constellations of states that minimize and maximize SU(2)-covariant Wigner negativity.  Comparisons are made to anticoherence, geometric entanglement, and $P$-representability; see Appendix \ref{sec:alternative_descriptions} for their descriptions.
\subsection{spin 1/2}
For the single qubit system ($j=1/2$) the stellar representation reduces to the Bloch sphere picture.  The Wigner functions of the spin-up and spin-down states along the $z$-axis (i.e.\ the standard computational basis states) are given by
\begin{equation}
    W_{\uparrow / \downarrow}(\theta,\phi) = \frac{1}{2} \pm \frac{\sqrt{3}}{2}\cos\theta
\end{equation}
Since all pure qubit states are trivially spin coherent states connected through rigid rotations, they all have the same Wigner negativity.  This has been calculated to be $\frac{1}{2} - \frac{1}{\sqrt{3}} \approx 0.077$ \cite{Davis_Wigner_2021, Arkhipov_2018}.

\subsection{spin 1}
\begin{figure}
    \centering
    \raisebox{0.25\height}{\includegraphics[width=0.2\textwidth]{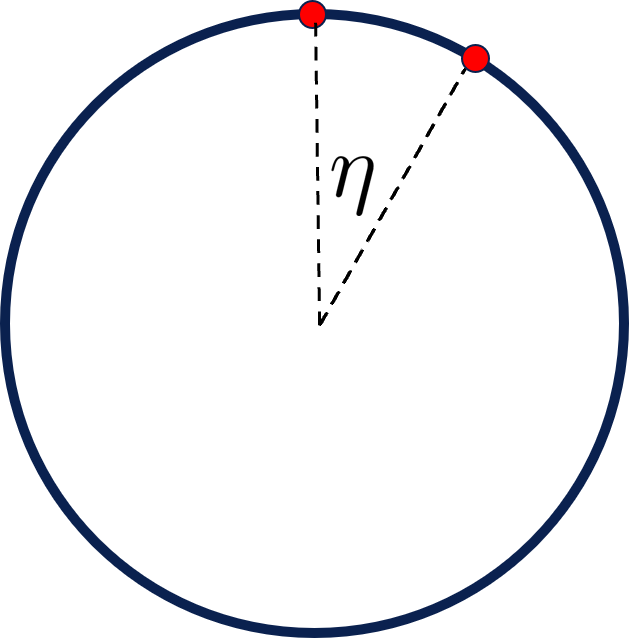}}
    \qquad \qquad
    \raisebox{0\height}{\includegraphics[width=0.45\textwidth]{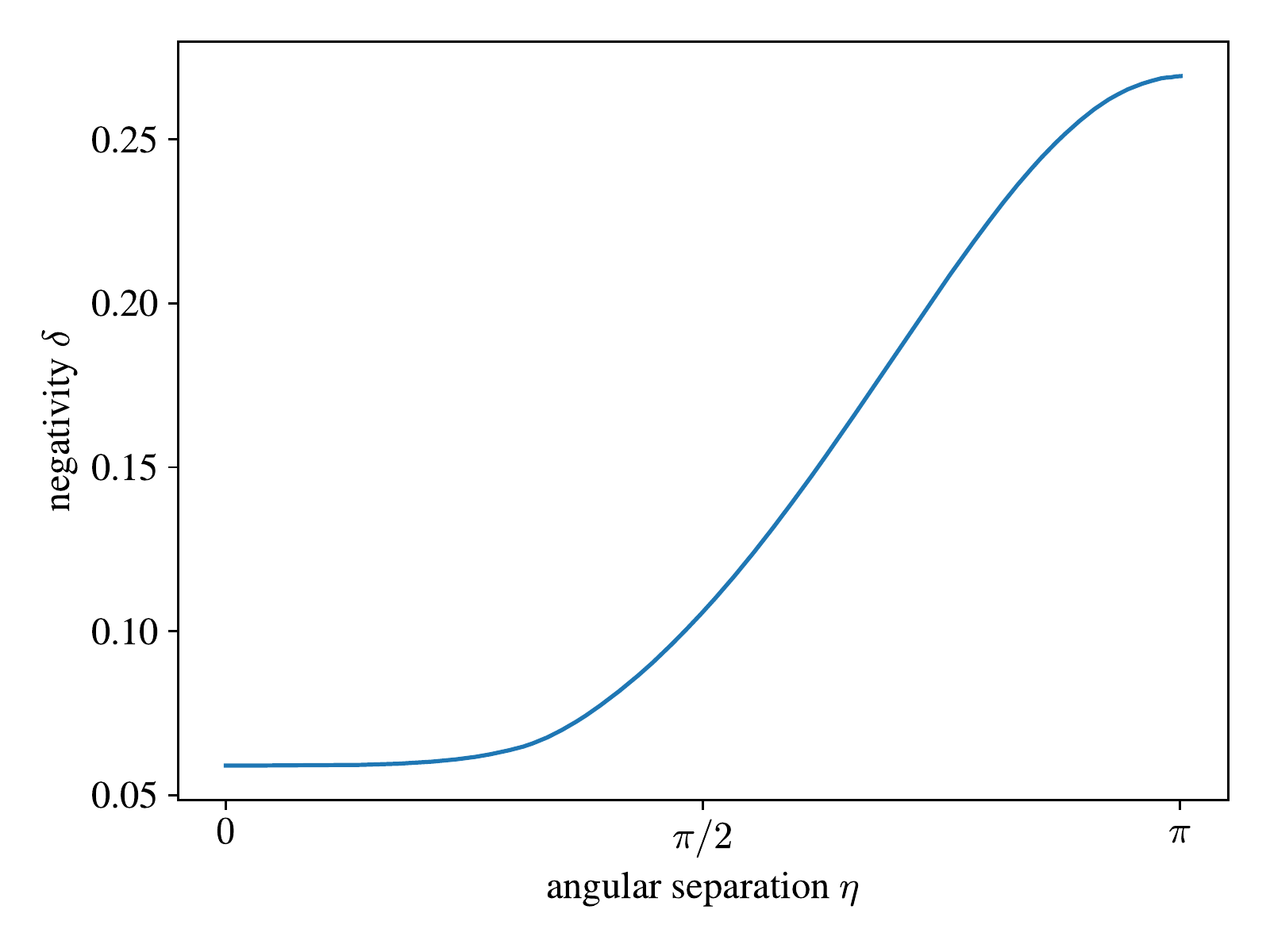}}
    \caption{Left: parameterization of all spin-$1$ states up to rigid rotational equivalence by the angular separation between the two stars.  Right: Wigner negativity as a function of polar separation.  The minimum is attained by the degenerate constellations (spin coherent states) while the maximum is obtained by the antipodal constellations (symmetric Bell states).}
    \label{fig:spin1-parameterization}
\end{figure}

Spin systems with $j=1$, equivalent to the symmetric subspace of two qubits, are characterized by two-point constellations.  Here we go into detail on how to characterize all such pure states; higher dimensions follow similarly.  Negativity is invariant under global rotation and so without loss of generality we fix one of the stars on the North pole.  The second is placed on the XZ plane with a polar separation $\eta$ relative to the former -- see Fig.\ \ref{fig:spin1-parameterization}.  Using Eq.\ \eqref{eq:symmetrizer}, this amounts to projecting the state $\ket{0}R_y(\eta)\ket{0}$ to the symmetric subspace followed by renormalization.
The family of $\eta$-parameterized states in the Dicke basis is calculated to be 
\begin{equation}
    \ket{\psi_\eta} = \frac{1}{\sqrt{1 + \cos^2\eta/2}} \bp{ \sqrt{2}\cos\frac{\eta}{2}\ket{1,1} + \sin\frac{\eta}{2}\ket{1,0} }.
\end{equation}
The corresponding family of Wigner functions is
\begin{equation}
    W_{\ket{\psi_\eta}}(\theta,\phi) = \frac{1}{1 + \cos^2\eta/2} \bs{ 2\cos^2\frac{\eta}{2} \, W_{\ket{1,1}}(\theta,\phi) + \sin^2\frac{\eta}{2} \, W_{\ket{1,0}}(\theta,\phi) + \frac{1}{\sqrt{2}}\sin\eta \, W_{\text{int}}(\theta,\phi) }
\end{equation}
where the two Dicke state terms, Eq.\ \eqref{eq:dicke-wigner}, are
\begin{equation}
    \begin{split}
        W_{\ket{1,1}}(\theta,\phi) &= \frac{1}{3}(1 - \sqrt{\frac{5}{8}}) + \sqrt{\frac{1}{2}} \cos\theta + \frac{1}{2} \sqrt{\frac{5}{2}}\cos^2\theta, \\
        W_{\ket{1,0}}(\theta,\phi) &= \frac{1}{3}(1 + \sqrt{\frac{5}{2}}) - \sqrt{\frac{5}{2}} \cos^2\theta
    \end{split}
\end{equation}
and the interference contribution is
\begin{equation}
     W_{\text{int}}(\theta,\phi) =  \sin\theta(1 + \sqrt{5}\cos\theta)\cos\phi.
\end{equation}
The Wigner negativity is then numerically computed for each polar separation $\eta$.  Every spin-1 pure state has a negativity value along the curve in Fig.\ \ref{fig:spin1-parameterization}.  This curve does not touch the horizontal axis and so, similar to the single qubit case, there is no pure state with vanishing negativity.  The states with minimal negativity are those with degenerate stars and correspond to the two-qubit spin coherent states.  We find, perhaps unsurprisingly, that the antipodal constellations maximize Wigner negativity.  This class of states is generated by one of the symmetric Bell states: $\frac{1}{\sqrt{2}}(\ket{01} + \ket{10}) = \ket{1,0}$ or $\frac{1}{\sqrt{2}}(\ket{00} + \ket{11}) = \frac{1}{\sqrt{2}}(\ket{1,1} + \ket{1,-1})$, together with their global rotations $\frac{1}{\sqrt{2}}(\ket{+-} + \ket{-+})$ or $\frac{1}{\sqrt{2}}(\ket{++}+\ket{--})$, etc.  The exact upper and lower bounds of the negativity are computed to be
\begin{equation}
\begin{split}
    \max_\eta \delta( W_\eta ) &= \frac{1}{3} \sqrt{\frac{2}{15} (55 + 17 \sqrt{10})} -1 \approx 0.2693 \\
    \min_\eta \delta( W_\eta ) &= \frac{4}{5 \sqrt{92 + 29 \sqrt{10}}} \approx 0.0590
\end{split}
\end{equation}
The alternative measures of nonclassicality are also minimized by degenerate configurations and maximized by antipodal configurations.

\subsection{spin 3/2}
\begin{figure}
    \centering
    \includegraphics[scale=0.65]{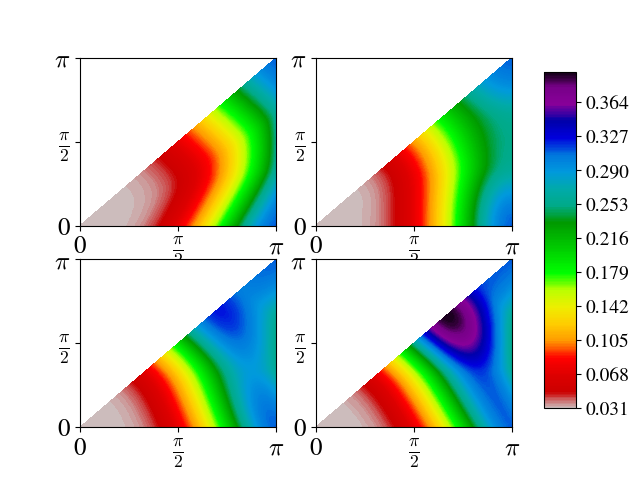}
    \caption{Cross-sections of the parameter space of three star constellations.  The value of each point represents the Wigner negativity of the spin-1 state associated with that constellation.  The axes represent the two polar angles $\vartheta_1$ and $\vartheta_2$, and each panel represents a different azimuthal separation $\varphi$: (a) $\varphi=0$, (b) $\varphi=\frac{\pi}{2}$, (c) $\varphi=\frac{3\pi}{4}$, (d) $\varphi=\pi$.  The most nonclassical constellation, $\vartheta_1=\vartheta_2=\frac{2}{3}$ and $\varphi=\pi$, is rotationally equivalent to the GHZ/$N00N$ state.}
    \label{fig:3qubit_parameter_negs}
\end{figure}

The number of free parameters needed to describe a pure $j=\frac{3}{2}$ state up to global rotations is three: two polar angles and one relative azimuthal angle.  Without loss of generality we place one star on the North pole and another on the XZ plane with polar angle $0 \leq \vartheta_1 \leq \pi$.  The final star has no constraints, having polar angle $0 \leq \vartheta_2 \leq \pi$ and azimuthal angle $0 \leq \varphi \leq \pi$ relative to the second star.  To avoid double counting the equivalent constellations $(\vartheta_1, \vartheta_2, \varphi)$ and $(\vartheta_2, \vartheta_1, \varphi)$ we also impose $\vartheta_2 \geq \vartheta_1$.  See Fig.\ \ref{fig:3qubit_parameter_negs}.  The family of states associated with these constellations is the projection of
\begin{equation}
    \ket{0}\otimes R_y(\vartheta_1)\ket{0}\otimes R_z(\varphi)R_y(\vartheta_2)\ket{0}
\end{equation}
to the symmetric subspace, given by
\begin{equation}
\begin{split}
    \ket{\psi_{\vartheta_1, \vartheta_2, \varphi}} &= N \Big[ 3\cos\frac{\vartheta_1}{2}\cos\frac{\vartheta_2}{2} \ket{\frac{3}{2},\frac{3}{2}} + \sqrt{3} \bp{\sin\frac{\vartheta_1}{2} \cos\frac{\vartheta_2}{2} + \cos\frac{\vartheta_1}{2}\sin\frac{\vartheta_2}{2} e^{i\varphi} } \ket{\frac{3}{2}, \frac{1}{2}} \\
    &\quad + \sqrt{3}\sin\frac{\vartheta_1}{2} \sin\frac{\vartheta_2}{2} e^{i\varphi} \ket{\frac{3}{2}, -\frac{1}{2}} \Big]
\end{split}
\end{equation}
up to normalization $N$.  The corresponding family of Wigner functions is similarly found via the generalized Weyl rule with respect to the $j=3/2$ SU(2) kernel. Fig.~\ref{fig:3qubit_parameter_negs} shows a selection of Wigner negativities.

Similar to the spin-1 system, the degenerate constellations have minimum but not vanishing negativity.  A representative constellation that maximizes negativity has $\vartheta_1 = \vartheta_2 = \frac{2\pi}{3}$ and $\varphi=\pi$, corresponding to the roots of unity along a great circle in the ZX plane.  This is rotationally equivalent to the GHZ/$N00N$ state quantized along the $z$-axis,
\begin{equation}
    \frac{1}{\sqrt{2}}(\ket{000} + \ket{111}) \simeq \frac{1}{\sqrt{2}}(\ket{3}\ket{0} + \ket{0}\ket{3}) \simeq \frac{1}{\sqrt{2}}(\ket{\frac{3}{2},\frac{3}{2}} + \ket{\frac{3}{2}, -\frac{3}{2}})
\end{equation}
respectively expressed in the computational basis, the two-mode occupation (Schwinger) representation \cite{schwinger2001angular}, and the Dicke basis.  This state has the same constellation but along the equator, and has widespread application in quantum information science and quantum metrology \cite{Dowling_2008, Bollinger_noon}.

By comparison, both anticoherence and $P$-representability similarly observe 3-cats as maximally nonclassical.  The geometric measure however is saturated by the antipodal constellation but with the North pole being two-star degenerate \cite{Aulbach_2010}.  This corresponds to the $W$ state $\frac{1}{\sqrt{3}}(\ket{001} + \ket{010} + \ket{100})$, which is incomparable to the spin cat state when restricted to LOCC operations in the qubit picture \cite{Dur_three_qubits_2002}.  The $W$ state has less negativity than the GHZ state.

\section{Higher spins}

The problem of determining the extremal quantum states for higher spin is technically more difficult, and we proceed primarily through numerical methods.  We evaluate the integral~\eqref{wigNeg} for the Wigner negativity numerically, seeking a precision of 5 digits in the final result.  To increase the likelihood that the constellation output is the true maximum of the Wigner negativity, we perform many iterations of the procedure, seeding the numerical optimization with different initial constellations selected randomly and uniformly across the sphere. 

In general the output will consist of several constellations that appear distinct. However, many of the constellations reached by the numerical solver will be related to one another by a rotation.  A simple technique to determine whether two constellations are definitively distinct is as follows. We first take the constellation and write the Cartesian vectors for each star in a matrix:
\begin{equation} 
A = \left[\vec{v}_1 ,  \, \vec{v}_2 ,\, \dots \, ,\vec{v}_n \right] \, .
\end{equation}
We then take the matrix $A A^T$ and compute its spectrum. If any two constellations have distinct spectra, then we can be certain the two constellations are also distinct. Because $A A^T$ is invariant if the individual vectors are acted upon by a rotation, if the spectra for two constellations are the same, then those constellations may be related by a rotation. 

Using these optimization and classification techniques, we perform a careful search to determine the constellation of maximum Wigner negativity. After a candidate has been determined, we run a secondary numerical optimization, this time constraining the search region to within $\pm 5\%$ of the previously determined values for the stars and working at machine precision. The negativity of the final candidate constellation is independently cross-checked using the methods in Refs.\ \cite{Koczor_continuous_2020, koczor_fast_computation_2020}.  Using many thousands of samplings of initial points, we check to ensure we find a global rather than a local maximum. 

\subsection{spin 2}

Spin systems with $j=2$ are characterized by four-star constellations. As usual, we fix the symmetries of the system by placing one star at the North pole, another along the $xz$ plane characterized just by its polar angle, and the remaining two are specified by both polar and azimuthal angles.  The remaining sections continue this pattern.  We find that the minimally Wigner-negative state corresponds to the constellation with all stars coincident at the North pole.  That is, up to rigid rotations the spin-2 coherent state is of minimal but not vanishing nonclassicality.

\begin{figure}
    \centering
    \includegraphics[width=0.45\textwidth]{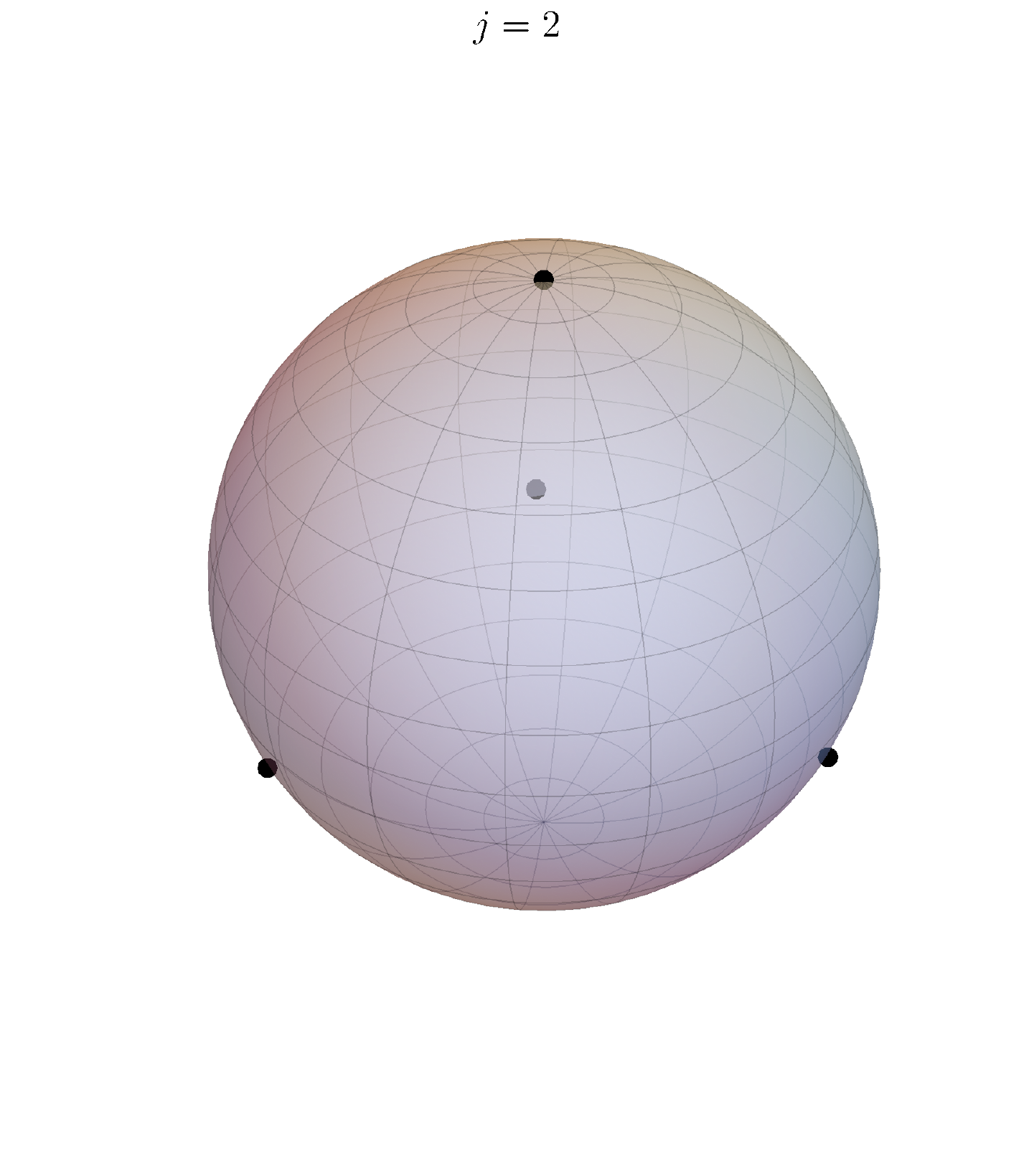}
    \quad \vspace{-1.5cm}
    \raisebox{0.3\height}{\includegraphics[width=0.45\textwidth]{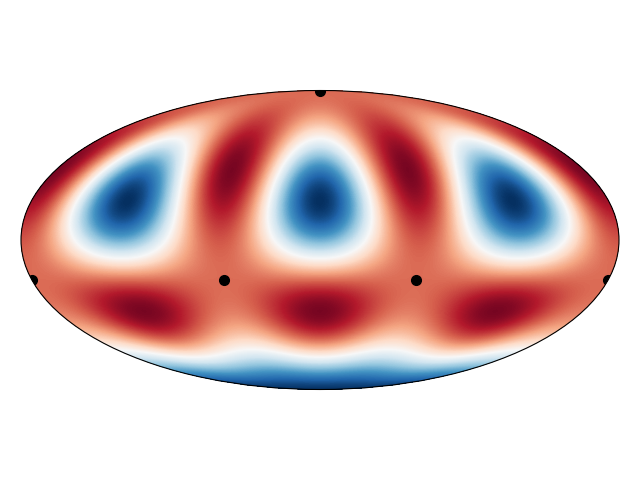}}
    \caption{Extremal state for spin 2. On the left we show the constellation of the extremal state, while on the right we show the Wigner function.}
    \label{spin2Max}
\end{figure}

We determine the state of maximal negativity to be the tetrahedron state, $\ket{\psi} = \frac{1}{\sqrt{3}}\ket{2,2} + \sqrt{\frac{2}{3}}\ket{2,-1}$.  The constellation consists of one point at the North pole and three points distributed with equiangular spacing along the azimuthal direction at a fixed polar angle of $\theta = 2 \arccos (1/\sqrt{3})$ (Fig.\ \ref{spin2Max}). It is interesting to note that all three comparative measures -- anticoherence, $P$-representability, and geometric entanglement -- also agree that the tetrahedron state is extremal for $j=2$.  This arrangement also solves several common spherical optimization problems such as the Thompson problem and the T\'oth problem \cite{Aulbach_2010}.  In addition to the perhaps expected cases of antipodal 2-qubit constellations and 3-qubit GHZ states, we will see this is the last instance where Wigner negativity agrees with any of the other nonclassicality measures.

\subsection{spin 5/2}

\begin{figure}
    \centering
    \includegraphics[width=0.45\textwidth]{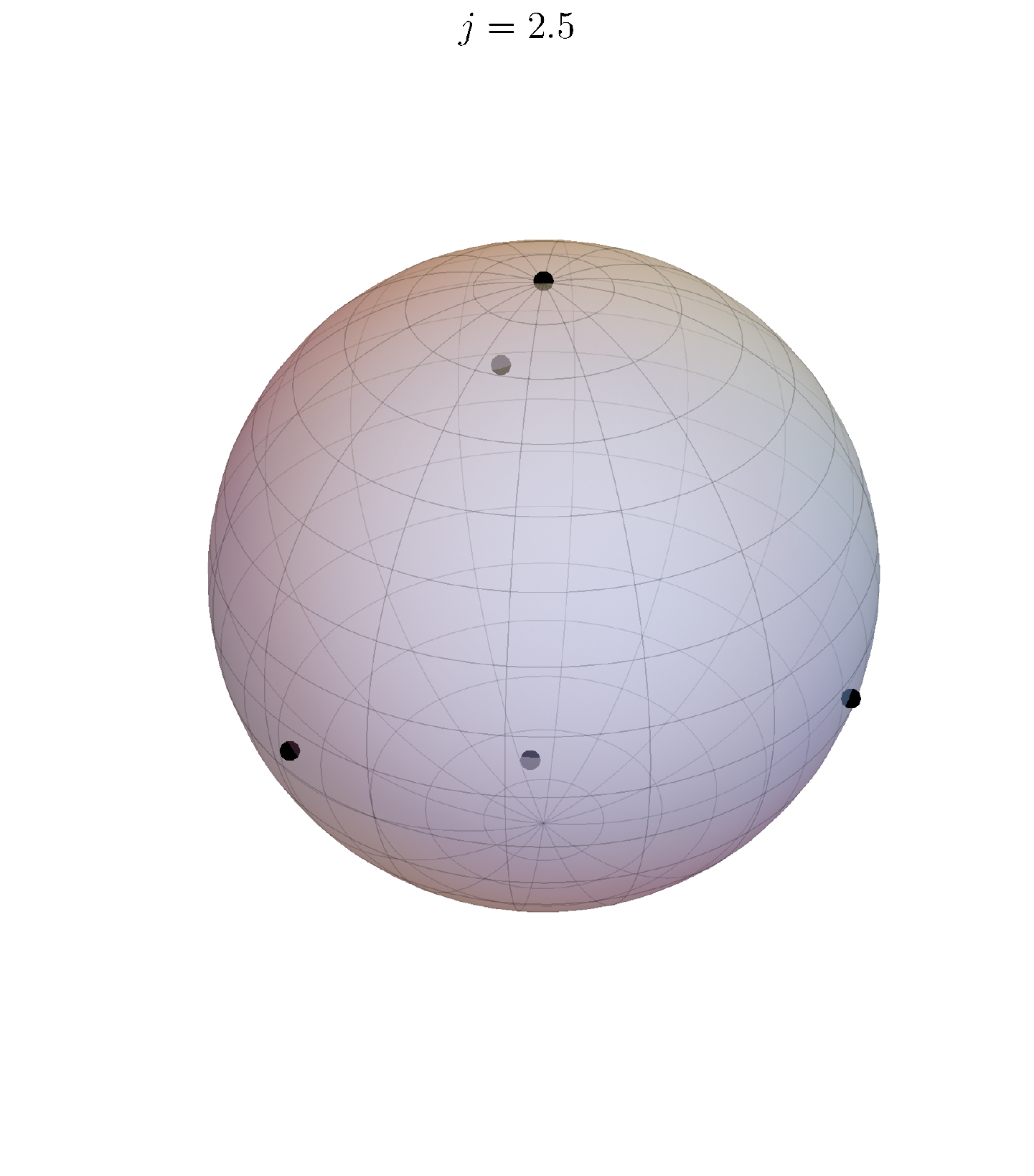}
    \quad \vspace{-1.5cm}
    \raisebox{0.3\height}{\includegraphics[width=0.45\textwidth]{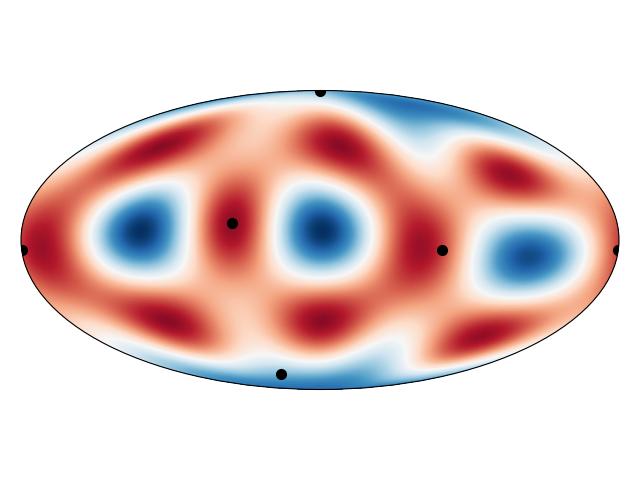}}
    \caption{Extremal state for spin $5/2$. Left is the associated constellation and right is the associated Wigner function.}
    \label{fig:spin2p5Max}
\end{figure}

Again, we have confirmed numerically that the spin coherent state is indeed the state of minimal Wigner negativity. For the maximal constellation we obtain the configuration shown in Fig.\ \ref{fig:spin2p5Max}. This differs from those found using other state nonclassicality measures. Compared to those alternative extremal states, it is not a particularly symmetric arrangement, and forms the beginning of a pattern of partially symmetric configurations as spin increases.

\begin{figure}
    \includegraphics[width=0.45\textwidth]{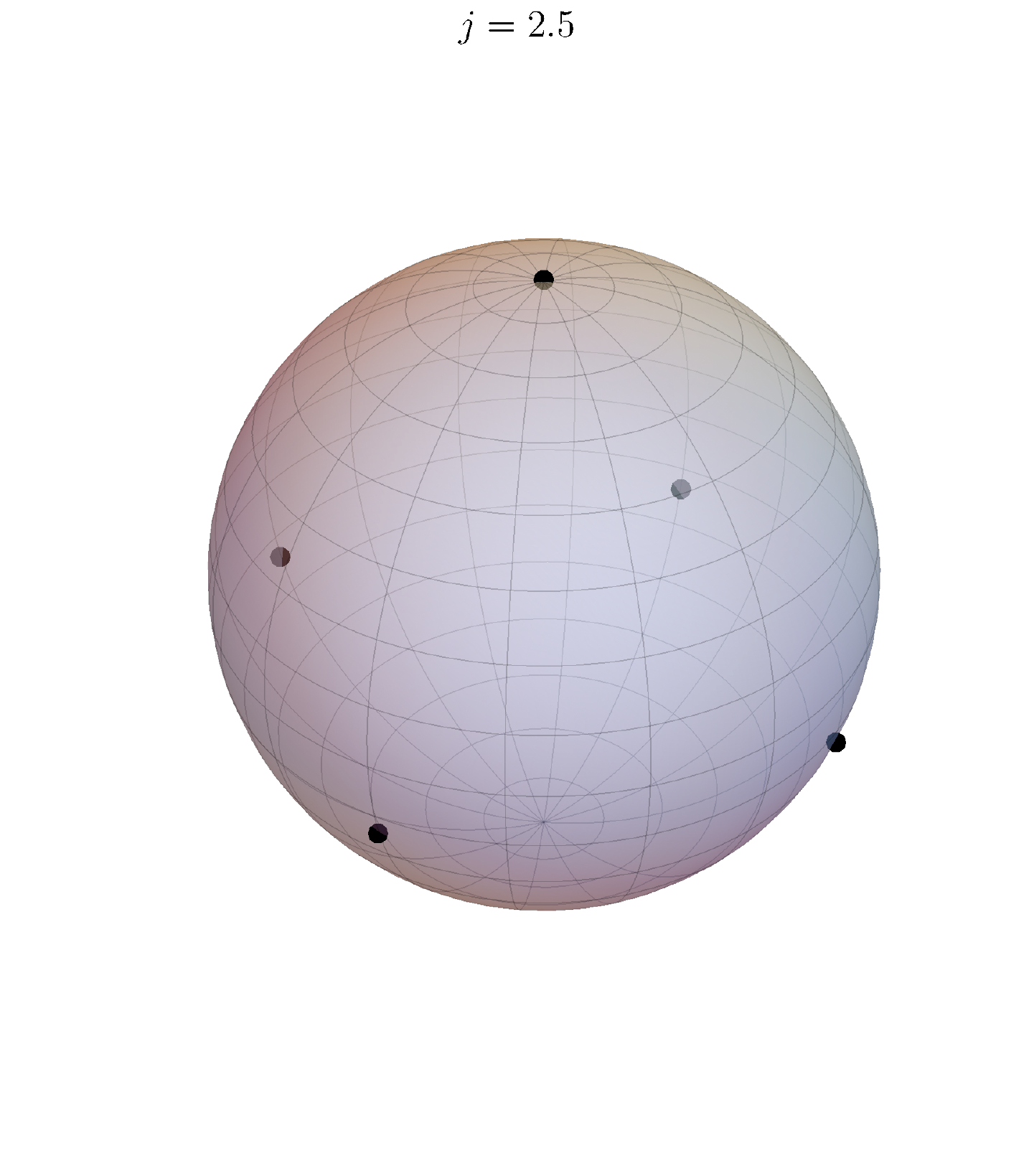}
    \quad \vspace{-1.5cm}
    \raisebox{0.3\height}{\includegraphics[width=0.45\textwidth]{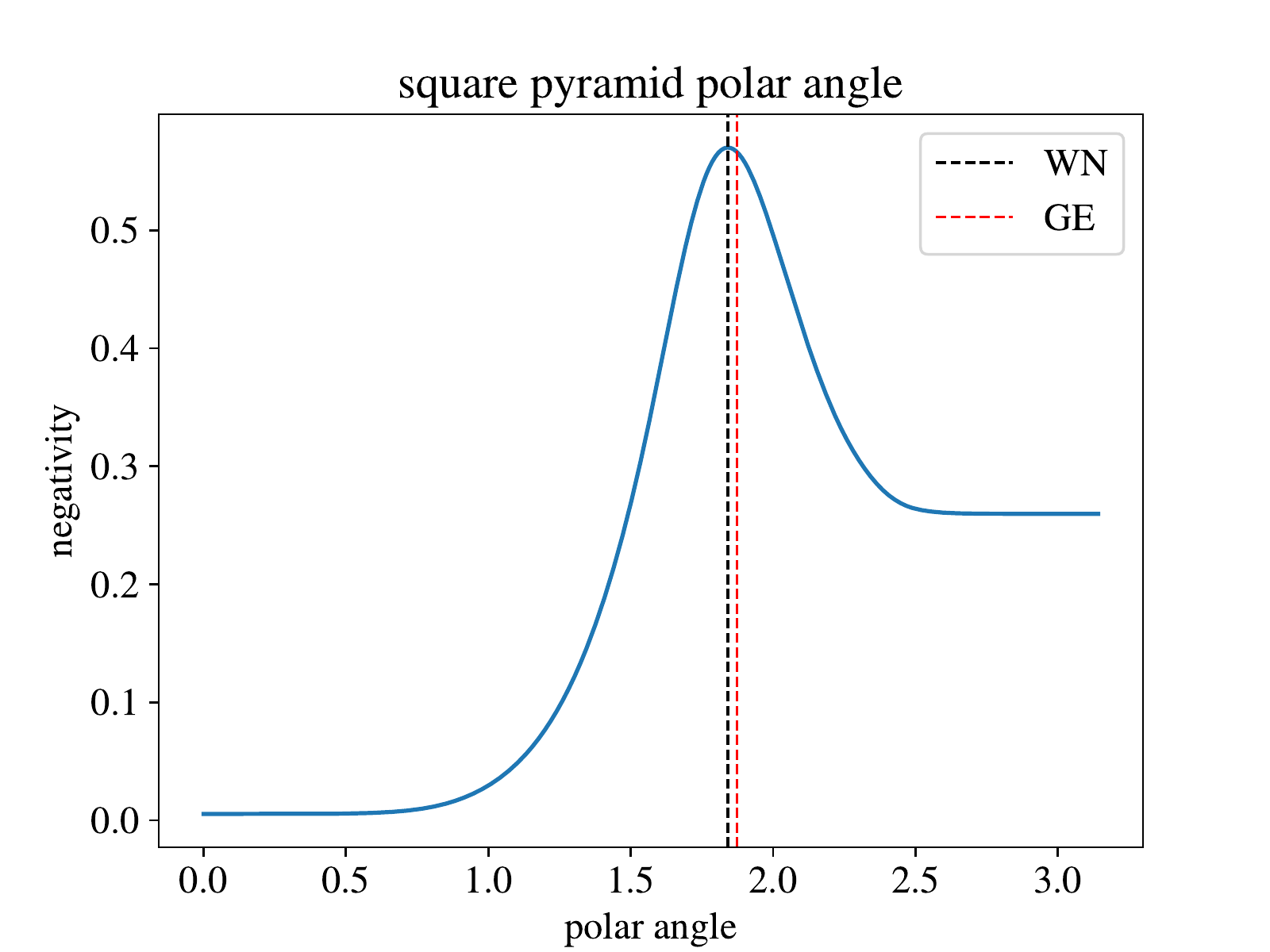}}
    \caption{Left: Constellation of a state that (locally) maximizes Wigner negativity. Right: Wigner negativities of the family of ``square pyramid" states parameterized by the polar angle of the pyramid base.  The stars in the base have azimuthal angles $\{0, \pi/2, \pi, 3\pi/2 \}$.  The state that maximizes geometric entanglement has a slightly larger pyramid height than the Wigner-maximal pyramid state.}
    \label{fig:spin2p5SubMax}
\end{figure}

We can make a further observation for $j = 5/2$: There is a second state with a local negativity maximum notably close in value to the global maximum.  The constellation for this second state is shown in Fig.\ \ref{fig:spin2p5SubMax}.  It is an embedded right square pyramid characterized by the base having polar angle $\theta \approx 1.841$. The Wigner negativity of this state is $\delta = 0.57015604$, which differs from maximum only at $\mathcal{O}(10^{-5})$. To ensure this is robust we  ran our numerical optimization scheme in a small neighbourhood of each constellation to higher precision -- again with independent verification from the methods described in Refs.\ \cite{Koczor_continuous_2020, koczor_fast_computation_2020} -- and found increasing numerical stability.  This pyramidal state is therefore an especially nonclassical spin-5/2 state, though not maximally nonclassical.  Interestingly, a regular square pyramid is also identified as the extremal constellation with respect to geometric entanglement \cite{Aulbach_2010}.  However, the two states are different as shown in Fig.\ \ref{fig:spin2p5SubMax}.  The maximally entangled pyramid has a base slightly further from the apex than the Wigner-maximal pyramid.  We also note that a pyramid state with base polar angle $\theta=\pi$ is equivalent to the 5-qubit $W$ state, which has a negativity $\approx 0.26$ as seen in Fig.~\ref{fig:spin2p5SubMax}.

\subsection{spin 3}

\begin{figure}
    \centering
    \includegraphics[width=0.45\textwidth]{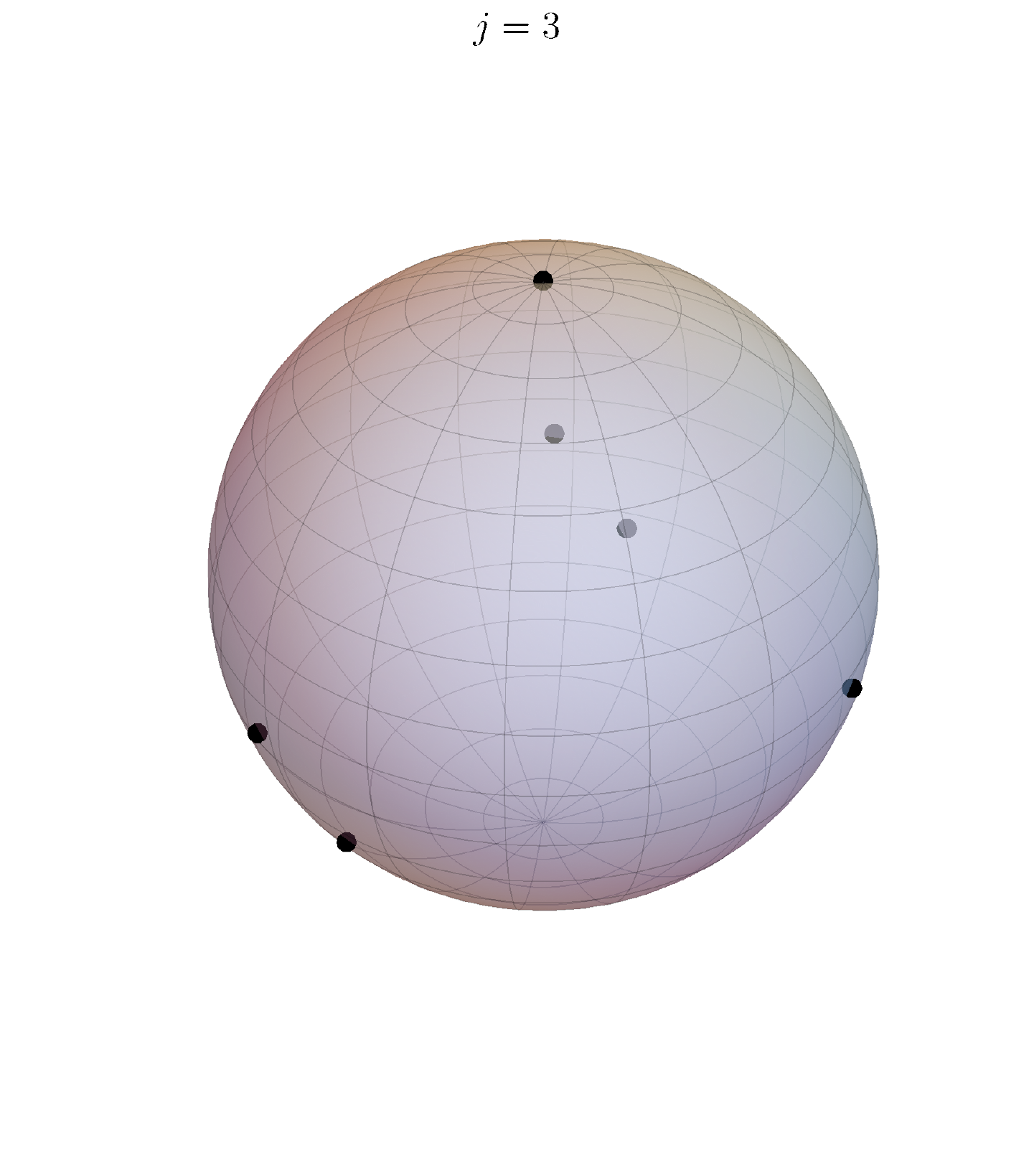}
    \quad
    \raisebox{0.3\height}{\includegraphics[width=0.45\textwidth]{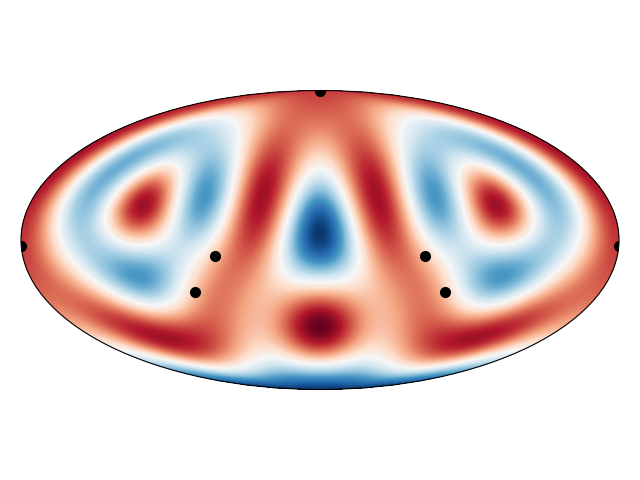}}
    \caption{Extremal state for spin $3$. On the left we show the constellation of the extremal state, while on the right we show the Wigner function.  This state is approximately 5\% more Wigner-negative than the octahedron state, which maximizes all the alternative notions of nonclassicality.}
    \label{fig:spin3Max}
\end{figure}

Spin systems with $j=3$ are characterized by six-star constellations.  The minimal constellation is again found to be the spin coherent state, and the maximal constellation is shown in Fig.\ \ref{fig:spin3Max}.  It is characterized by four co-planar points that together form a rectangle, along with the star at the North pole and another displaced along the arc $\varphi = 0$.  The case of spin 3 is notable because a different state, $\frac{1}{\sqrt{2}}(\ket{3,2} + \ket{3,-2})$, simultaneously maximizes the alternative measures.  This is the so-called octahedron state, and is characterized by a constellation with stars along the vertices of an embedded octahedron.  With an \textit{a priori} presumption that nonclassicality correlates with constellation symmetry, one would perhaps expect the Wigner case to follow suit because the octahedron is the next available Platonic solid as the number of stars increase.  Yet the most Wigner-negative state identified here is approximately 5\% more negative than the octahedron state, a significantly higher gap than the pyramidal runner-up in the five qubit system.

\begin{figure}
    \centering
    \includegraphics[width=0.4\textwidth]{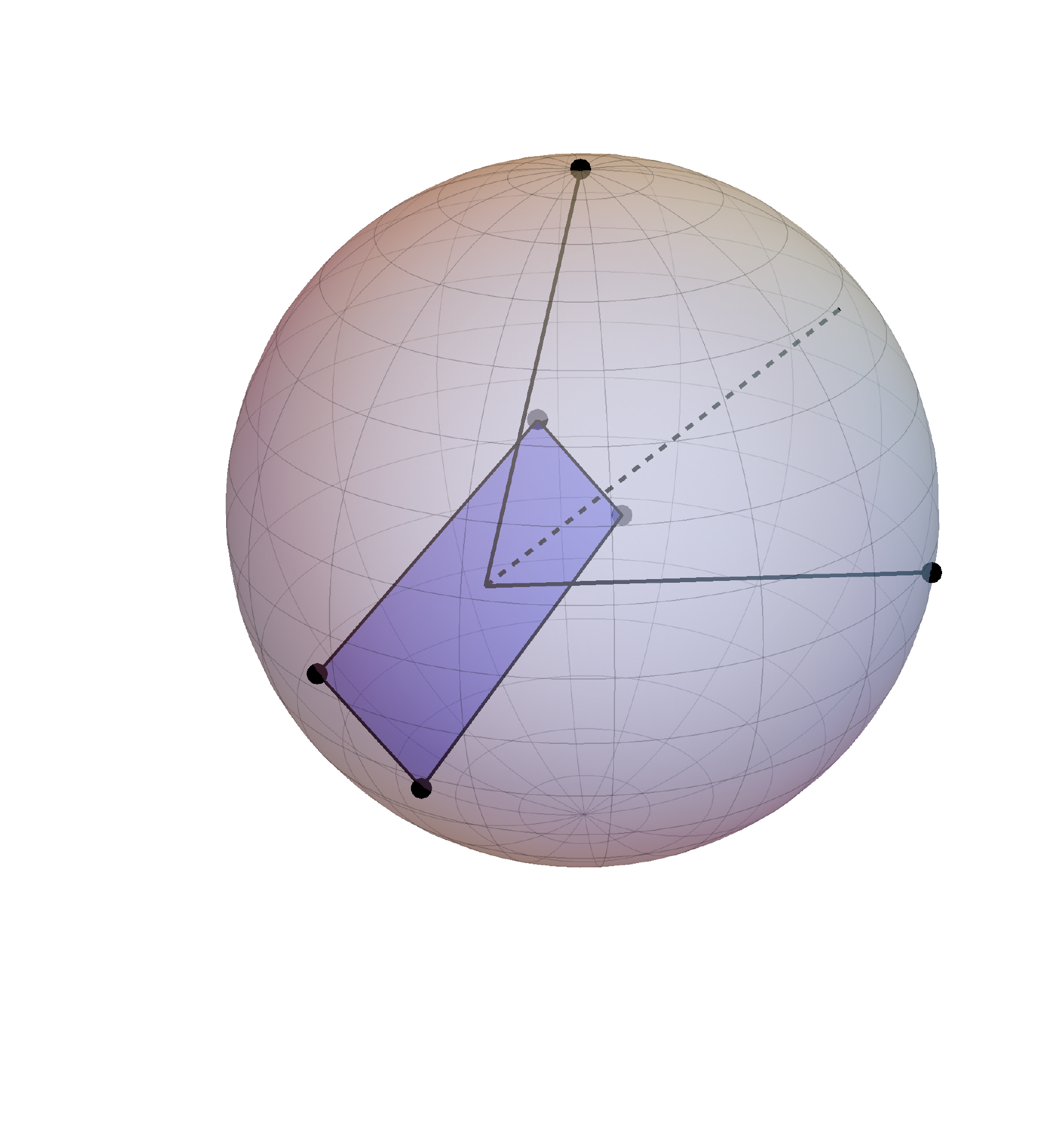}
    \qquad \vspace{-1cm}
    \includegraphics[width=0.4\textwidth]{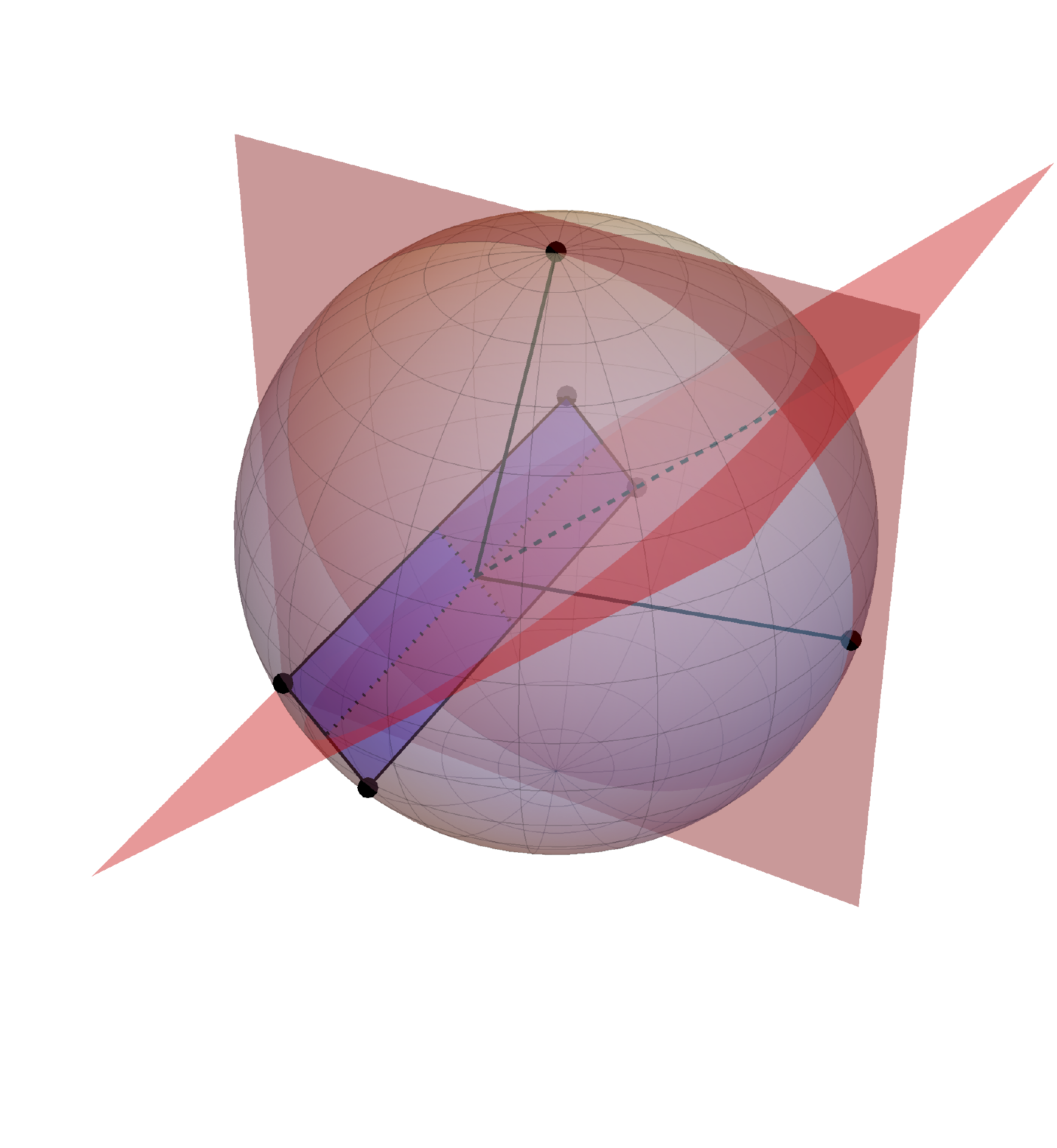}
    \caption{Demonstrating $C_{2\text{v}}$ point symmetry of the maximal spin 3 constellation.  Left:  $\pi$-rotation about dotted line.  Right: Two mirror planes parallel to the axis of rotation.}
    \label{fig:spin3planes}
\end{figure}

\begin{figure}
    \centering
    \includegraphics[width=0.3\textwidth]{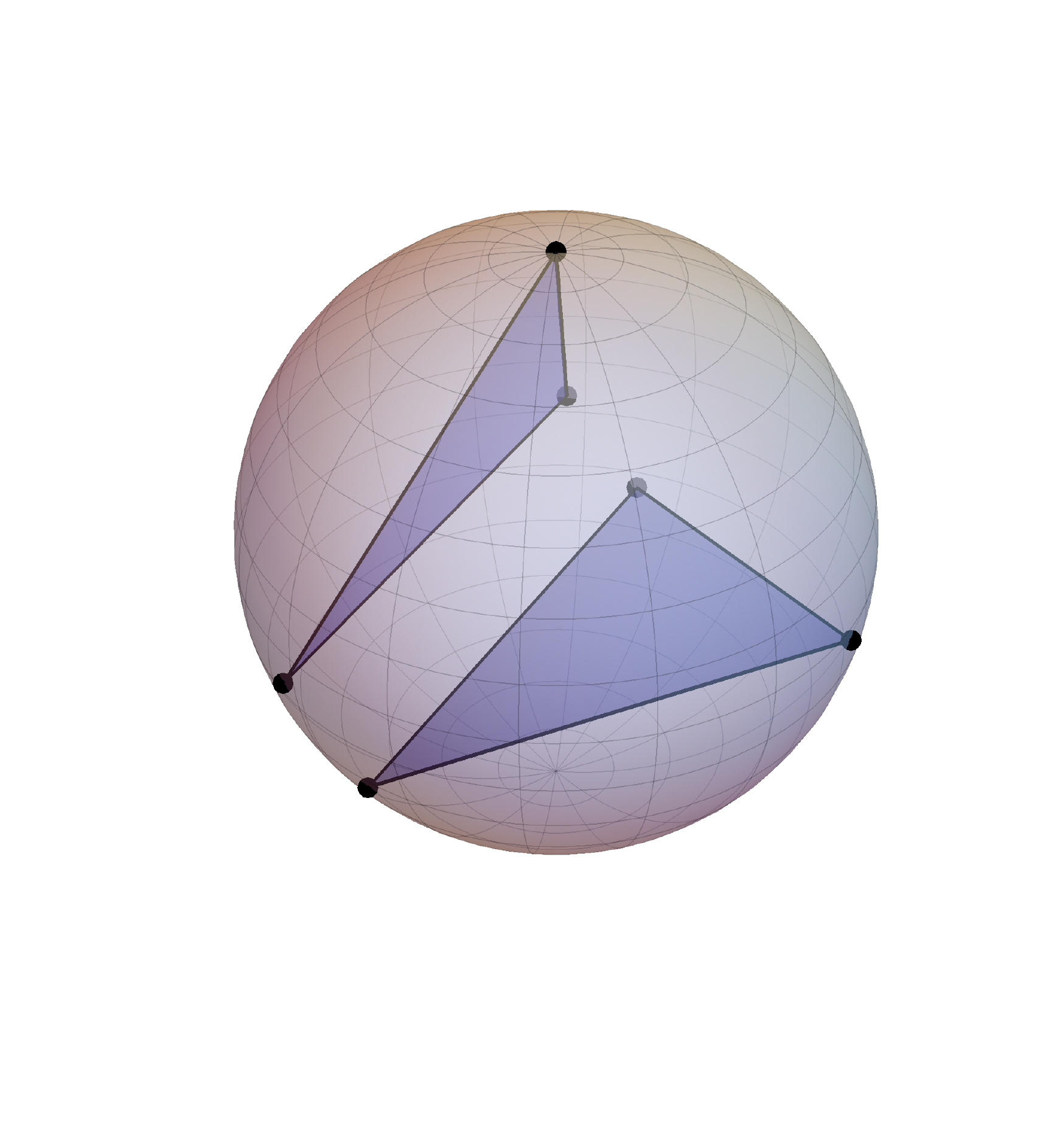}
    \includegraphics[width=0.3\textwidth]{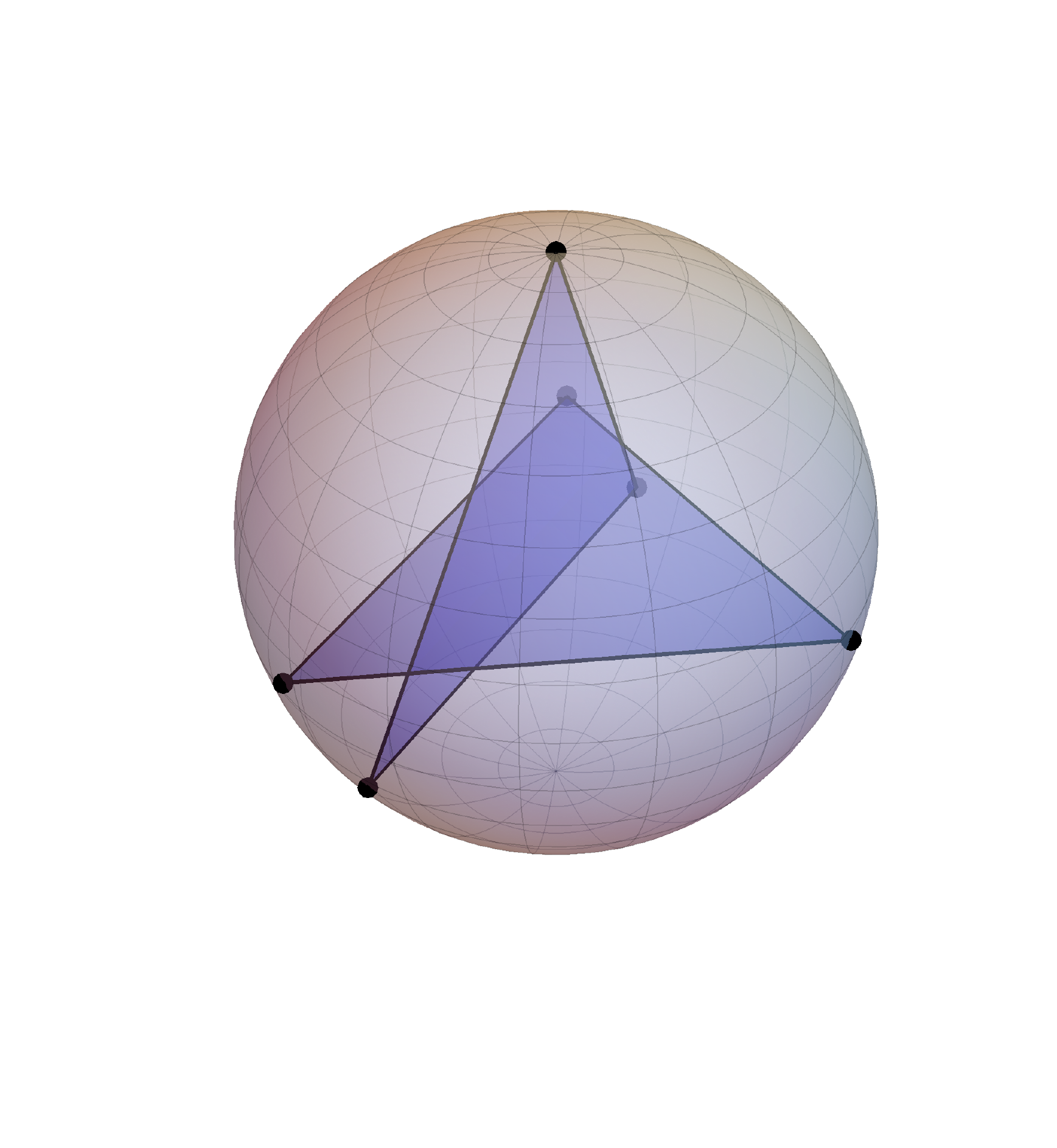}
    \includegraphics[width=0.3\textwidth]{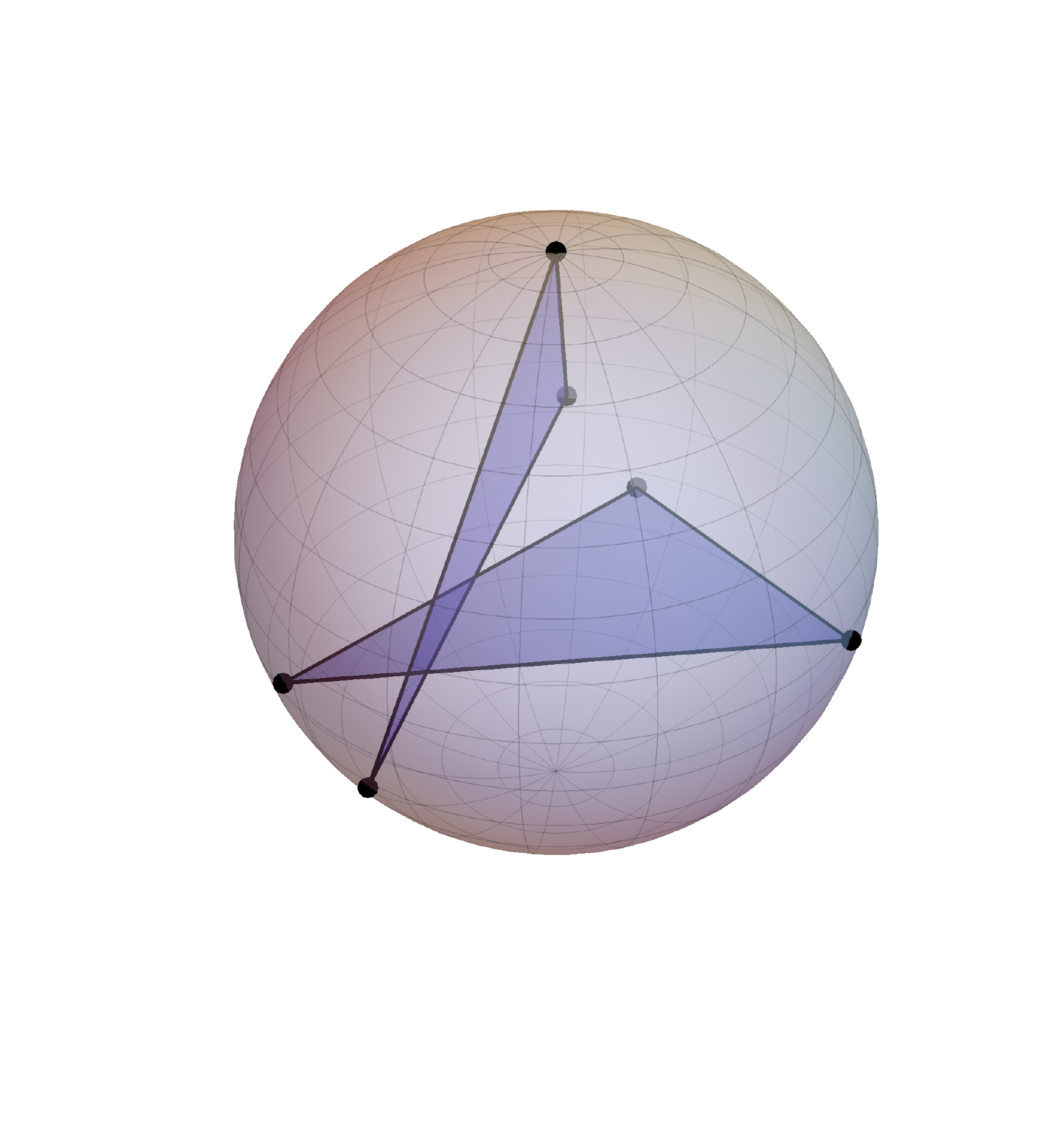}
    \includegraphics[width=0.3\textwidth]{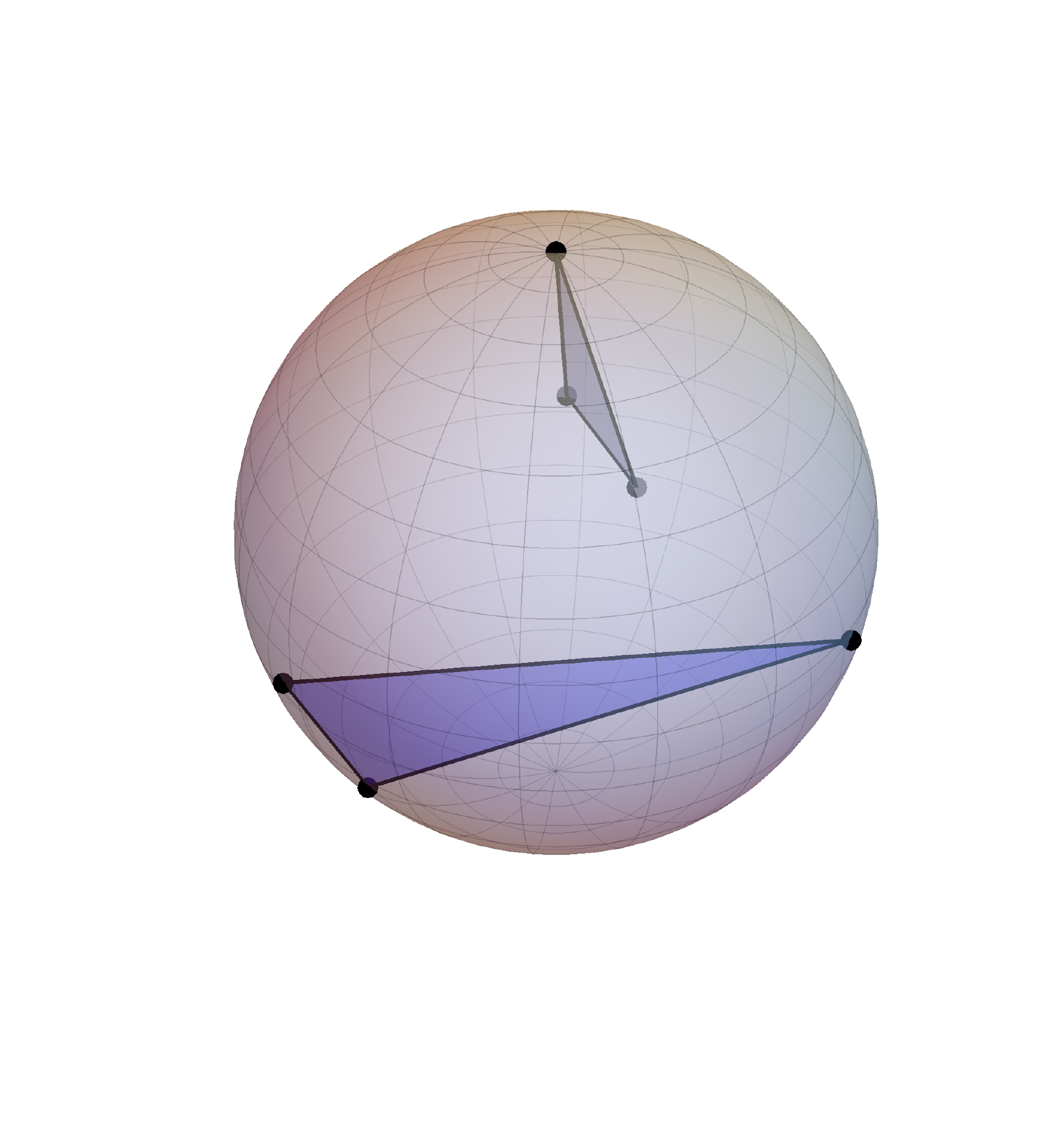}
    \includegraphics[width=0.3\textwidth]{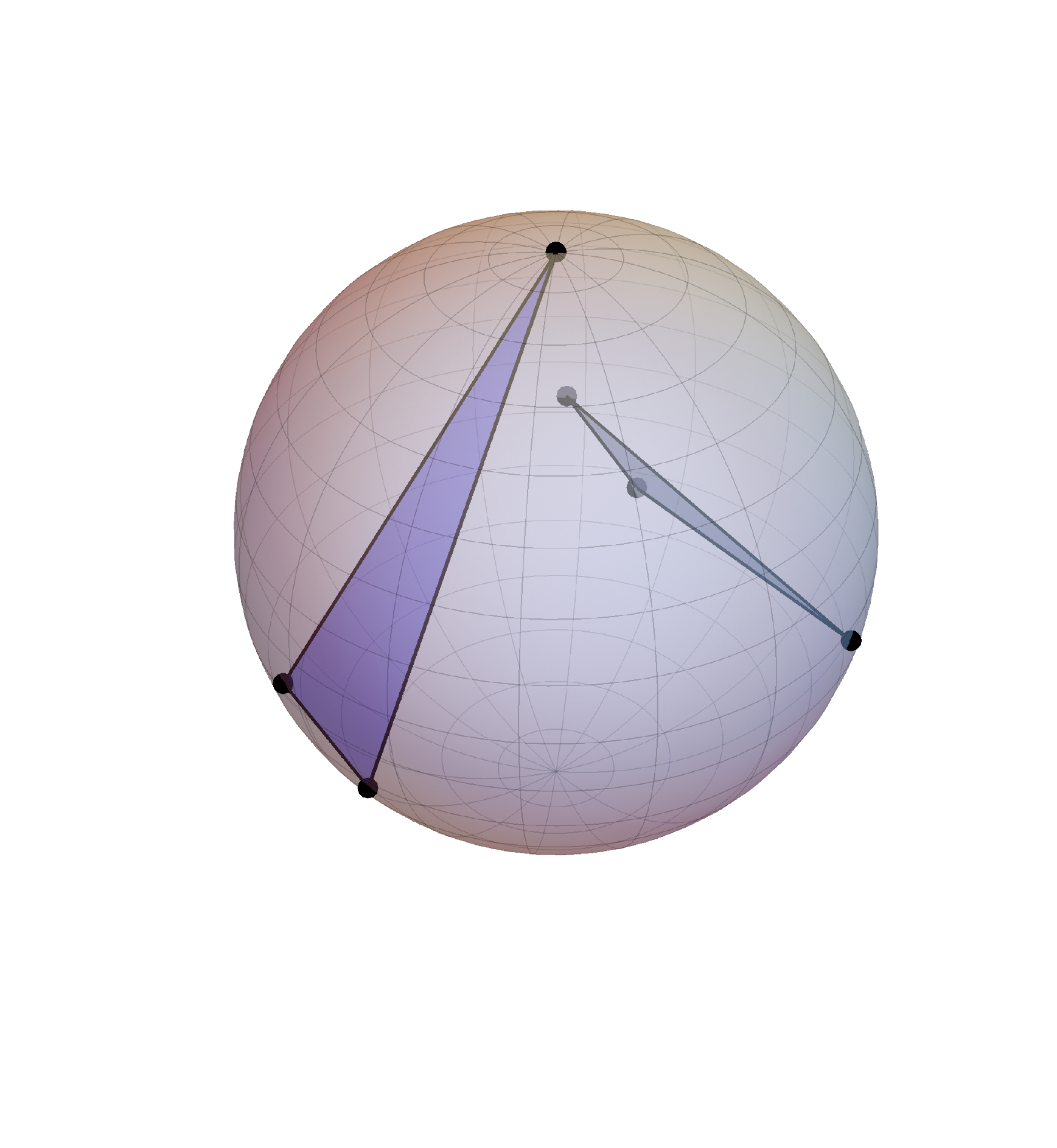}
    \includegraphics[width=0.3\textwidth]{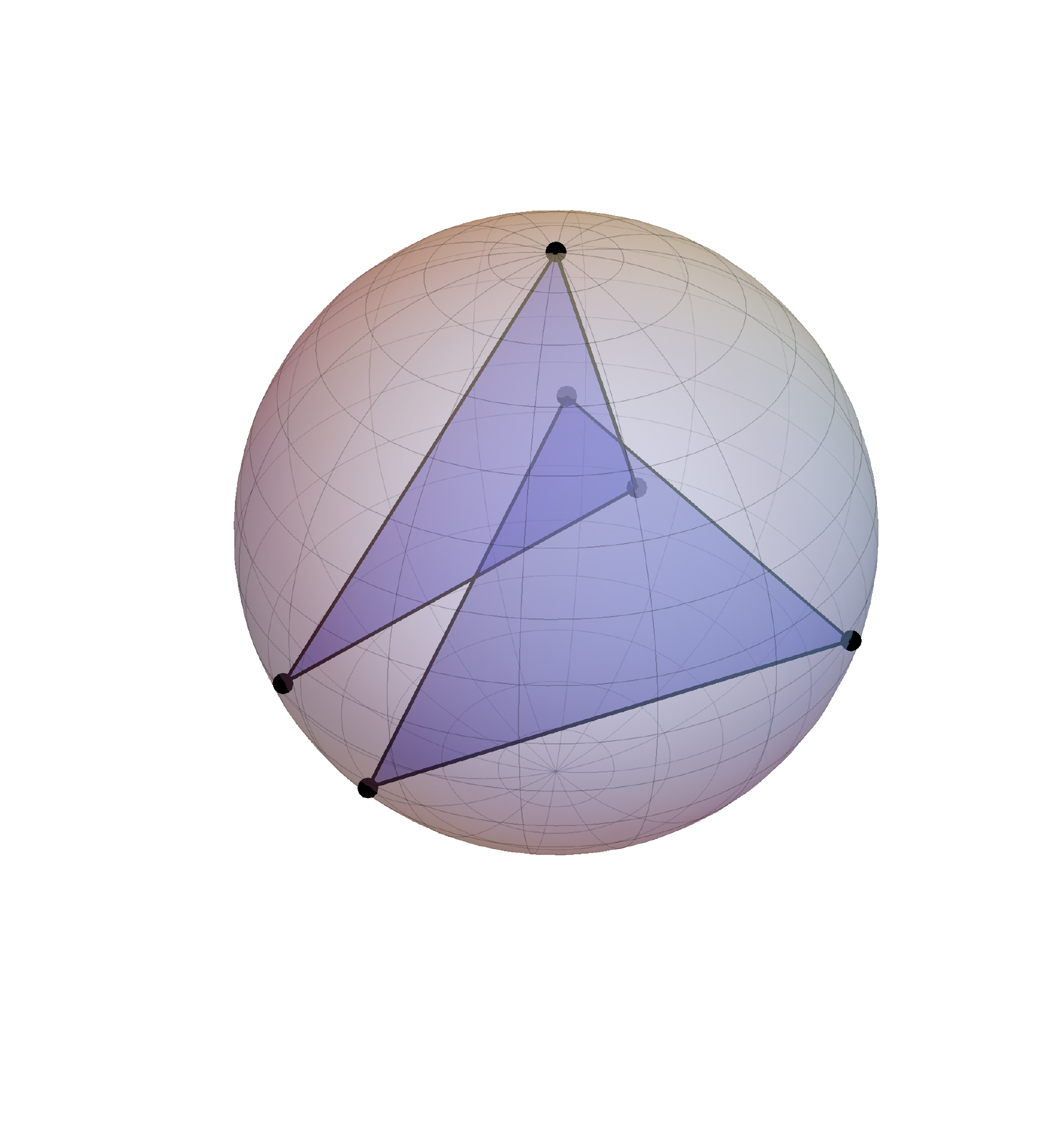}
    \vspace{-1cm}
    \caption{Bi-partitions of the spin-3 maximal state into identical triangles.  In the top-left and top-middle plots the pair of triangles are isosceles; the others are scalene.  In every case the triangles are identical --- their angles and edge lengths are equal within working precision.  In maintaining a common perspective this may be less apparent in some cases.}
    \label{fig:spin3triangles}
\end{figure}

Nonetheless, the $j=3$ constellation is not without symmetry.  As shown in Fig.~\ref{fig:spin3planes}, the configuration exhibits $C_{2\text{v}}$ point group invariance.  The stars can also be bi-partitioned into two identical triangles in six different ways; see Fig.~\ref{fig:spin3triangles}.  The first two partitions, shown in the upper-left and upper-middle plots of Fig.~\ref{fig:spin3triangles}, yield a pair of isosceles triangles.  The remaining partitions are made of pairs of identical scalene triangles.  The octahedron constellation by comparison can be partitioned into two identical triangles in ten different ways, which is the maximal number of such partitions.  For reference, in the upper-left partition the triangles have angles $53.9^\circ$, $53.9^\circ$, and $72.2^\circ$ with edge lengths $1.51$, $1.51$, and $1.78$.  The upper-middle triangles have angles $58.3^\circ$, $58.3^\circ$, and $63.4^\circ$ with edge lengths $1.69$, $1.69$, and $1.77$.

\begin{figure}
    \centering
    \includegraphics[width=0.45\textwidth]{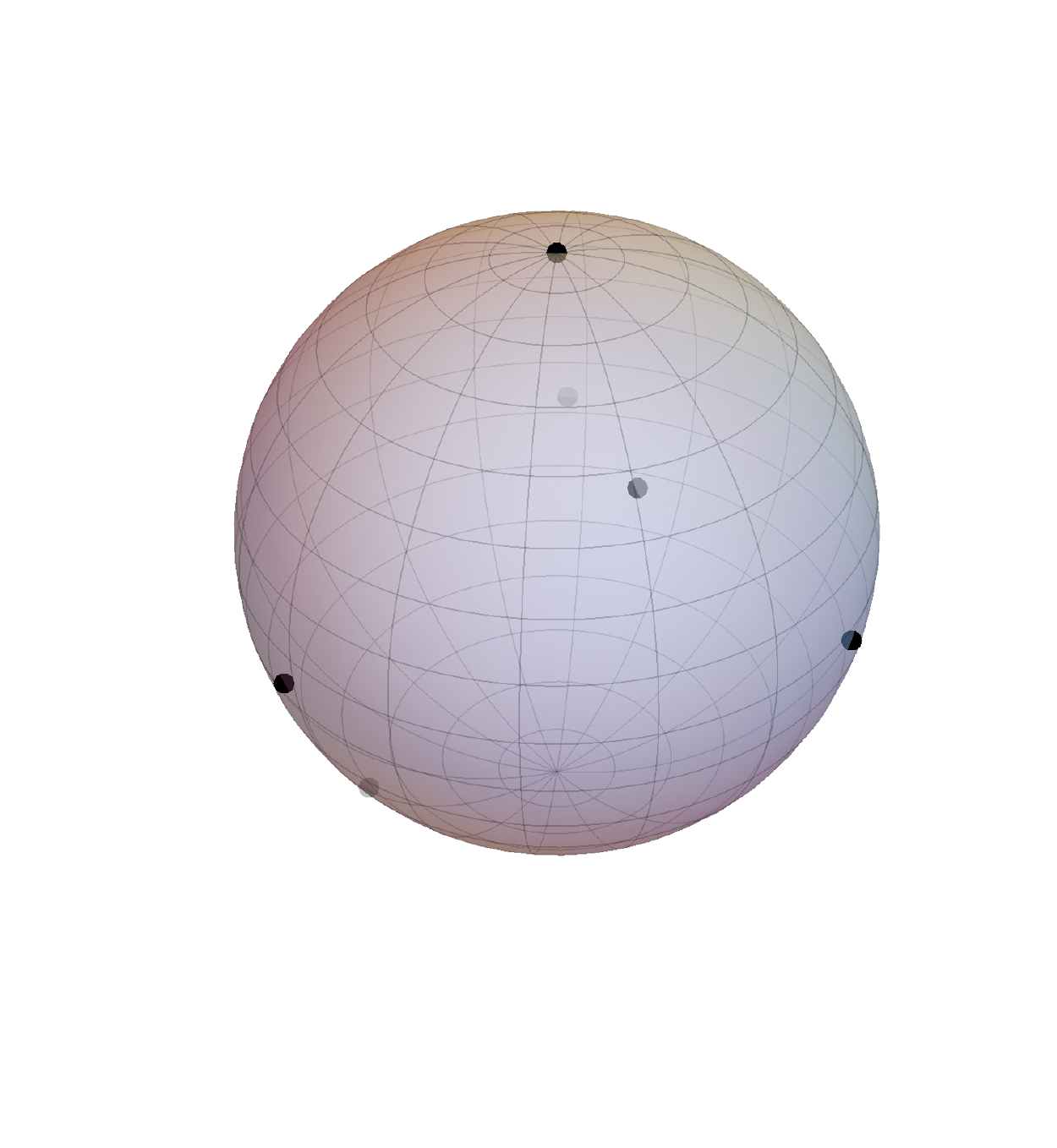}
    \quad \vspace{-2cm}
    \includegraphics[width=0.45\textwidth]{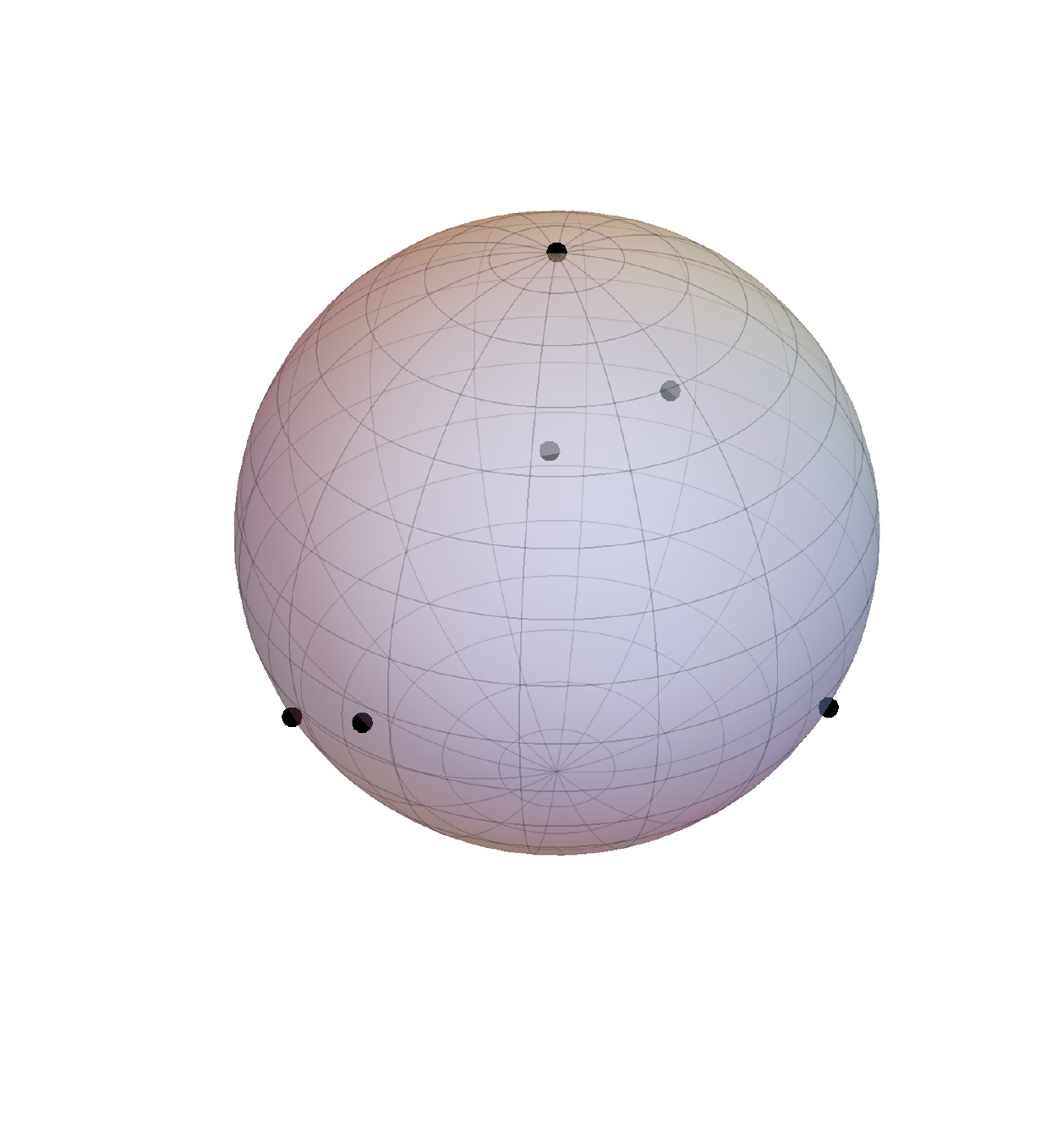}
    \caption{Left: maximal spin-3 state with four highlighted stars that approximate a regular tetrahedron.  Right: constrained optimization after snapping the four points to a regular tetrahedron. The right has a Wigner negativity within $2\%$ of the left. }
    \label{fig:spin3ApproxTet}
\end{figure}

We also mention another symmetric and highly Wigner-negative constellation.  If four of the six stars are required to form a tetrahedron, with the other two varied over the numerical optimization, we find the constellation shown on the right of Fig.~\ref{fig:spin3ApproxTet}.  This state appears to be a deformed version of the true maximum, shown on the left.  The forced-tetrahedron state has a Wigner negativity of $0.64521$, bringing it within $2\%$ of the true maximum.  This configuration also surpasses the octahedron state in Wigner negativity.

\subsection{spin 7/2}

\begin{figure}
    \centering
    \includegraphics[width=0.45\textwidth]{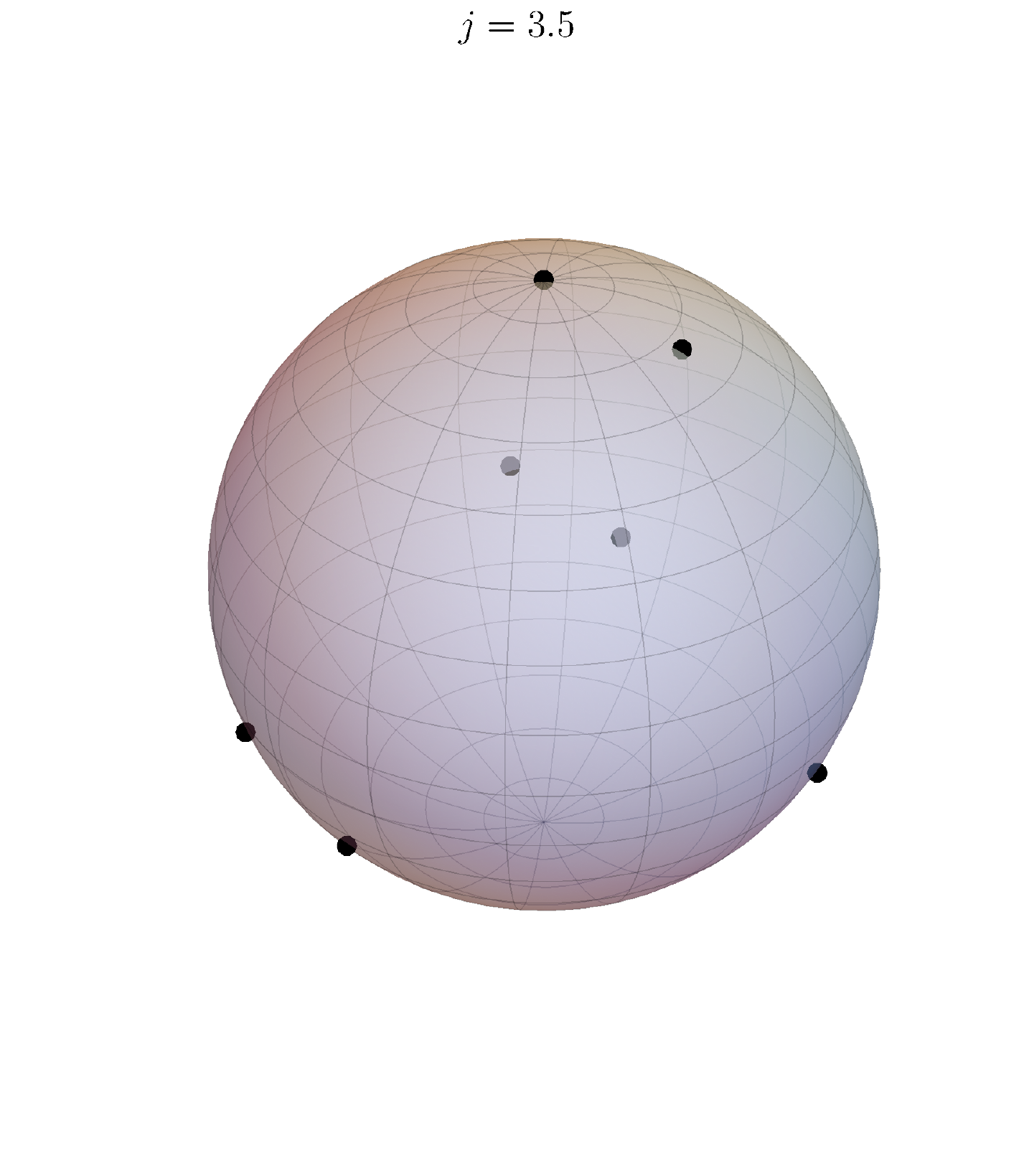}
    \quad \vspace{-1cm}
    \raisebox{0.3\height}{\includegraphics[width=0.45\textwidth]{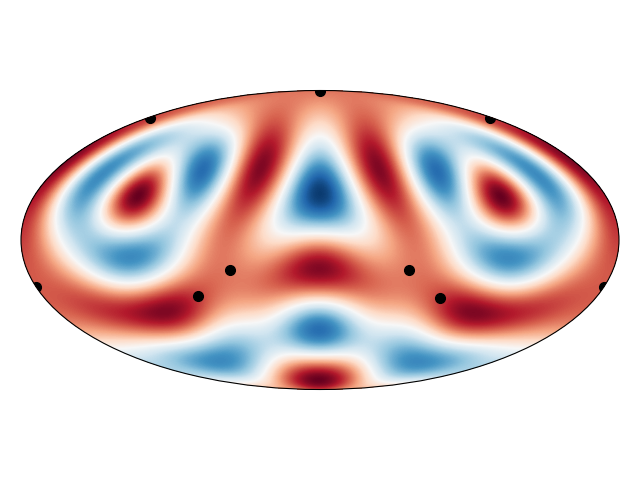}}
    \caption{Maximal spin-$\frac{7}{2}$.  Left is the constellation and right is the Wigner function.}
    \label{fig:spin3p5Max}
\end{figure}

\begin{figure}
    \centering
    \includegraphics[width=\textwidth]{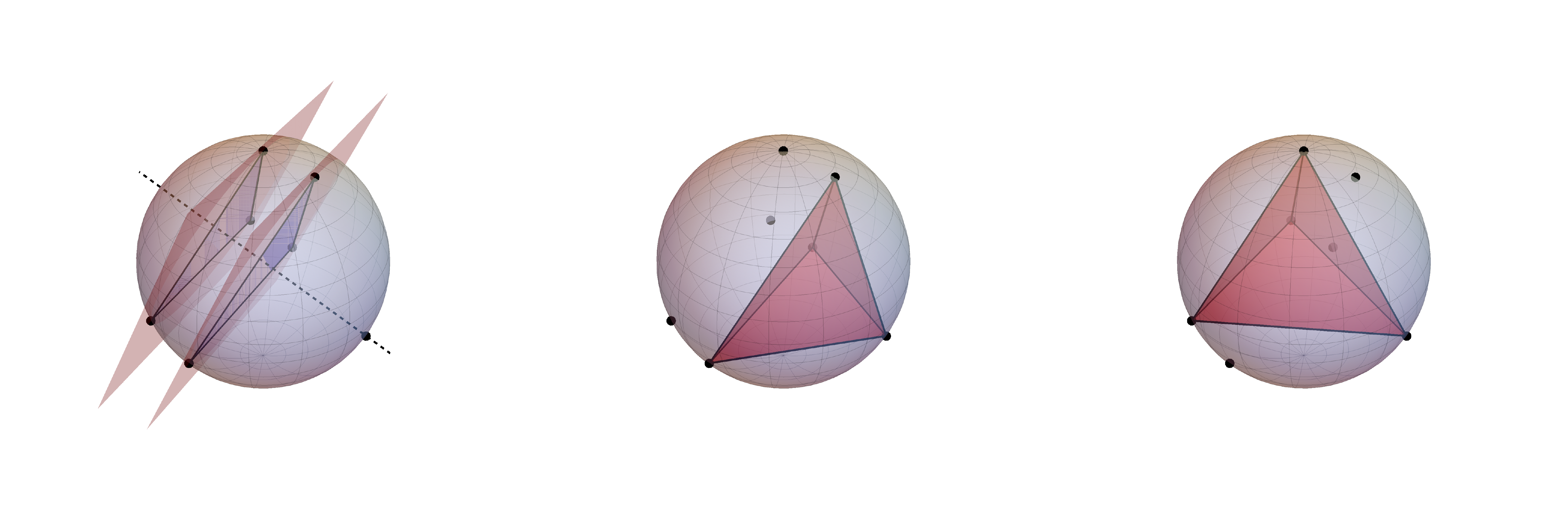}
    \vspace{-1.5cm}
    \caption{Left: parallel planes extended by two equilateral triangles within the maximal spin-$\frac{7}{2}$ constellation.  The dotted line is a diameter passing through the centroid of the triangles and the remaining point.  Middle and Right: irregular tetrahedra within the constellation, with the right closely approximating a regular tetrahedron.}
    \label{fig:7qubits_combined}
\end{figure}

\begin{figure}
    \centering
    \includegraphics[scale=0.5]{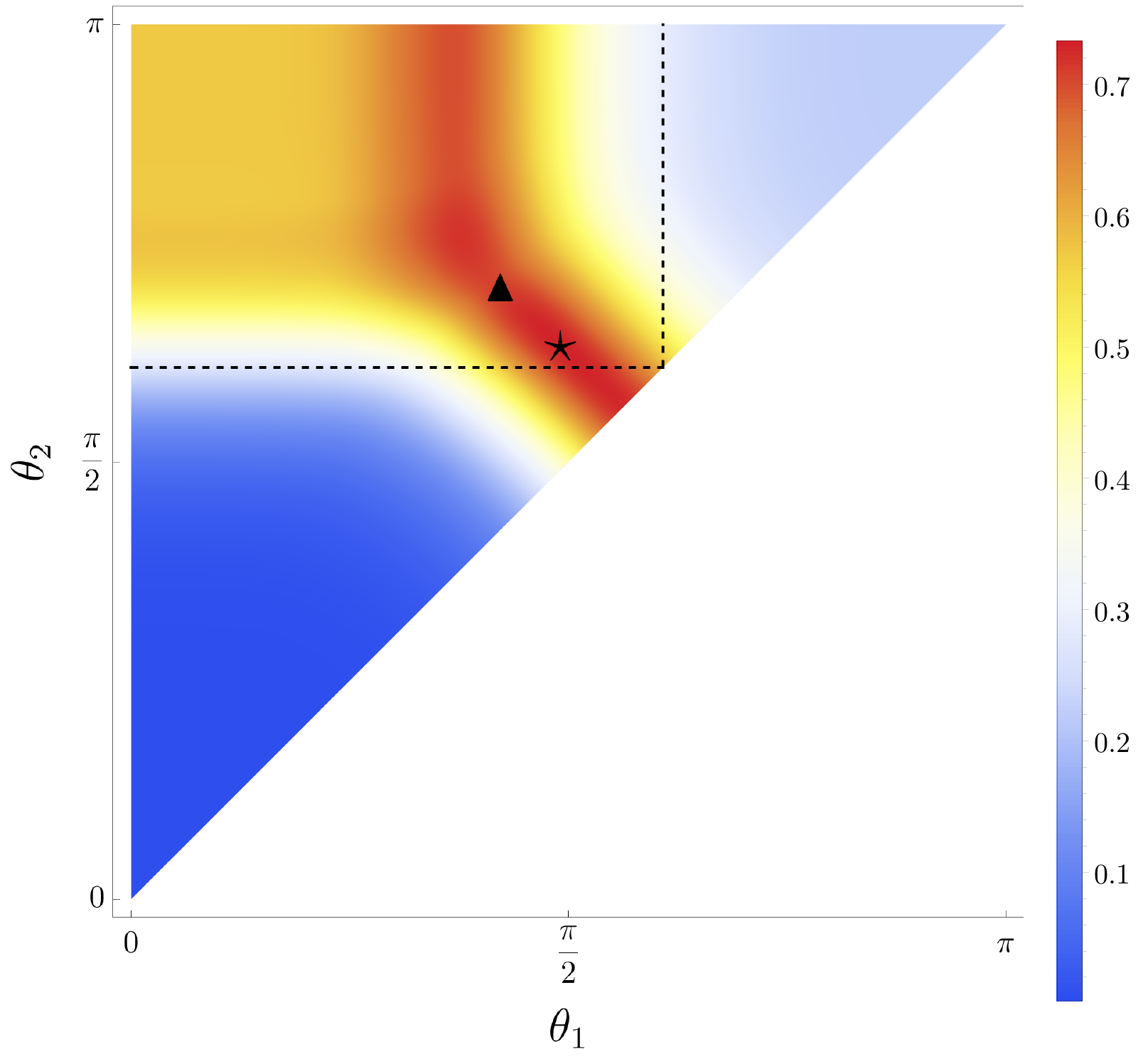}
    \caption{Wigner negativity landscape of ``two triangles + pole" constellations.  Parameters are the polar angles of each triangle with respect to the axis of rotational symmetry. The dashed lines indicate when one of the triangles forms a tetrahedron with the pole. The star indicates the Wigner extremal state, while the triangle indicates the anticoherence extremal state.}
    \label{fig:7qubit-triangles-param}
\end{figure}

Spin systems with $j=7/2$ correspond to seven stars.  The maximal constellation is characterized by two parallel equilateral triangles with matching orientation, together with a star along the diameter going through the centroids of the triangles; see Figs.~\ref{fig:spin3p5Max} and \ref{fig:7qubits_combined}.  Using a similar argument as the six star system, such constellations can be seen to have $C_{3\text{v}}$ point group symmetry.  The maximally anticoherent state is also of this ``two triangles + pole" form but with different heights of the triangles along the rotational axis \cite{bjork_kings_2015}.  In particular, the two triangles in the Wigner case are significantly closer to each other: $\Delta d_{\text{Wigner}} \approx 0.43$ and $\Delta d_{\text{ac}} \approx 0.82$ as measured by their axial distance along the diameter.  See Fig.~\ref{fig:7qubit-triangles-param} for a comparison between all such constellations as parameterized by the polar angles of the two triangles (polar with respect to the axis pointing towards the 7th star).  By comparison, the maximal constellations of both geometric entanglement and $P$-representability are described by a pentagon along the equator together with a star on each pole \cite{Aulbach_2010,Giraud_2010}.

\begin{figure}
    \centering
    \includegraphics[scale=0.5]{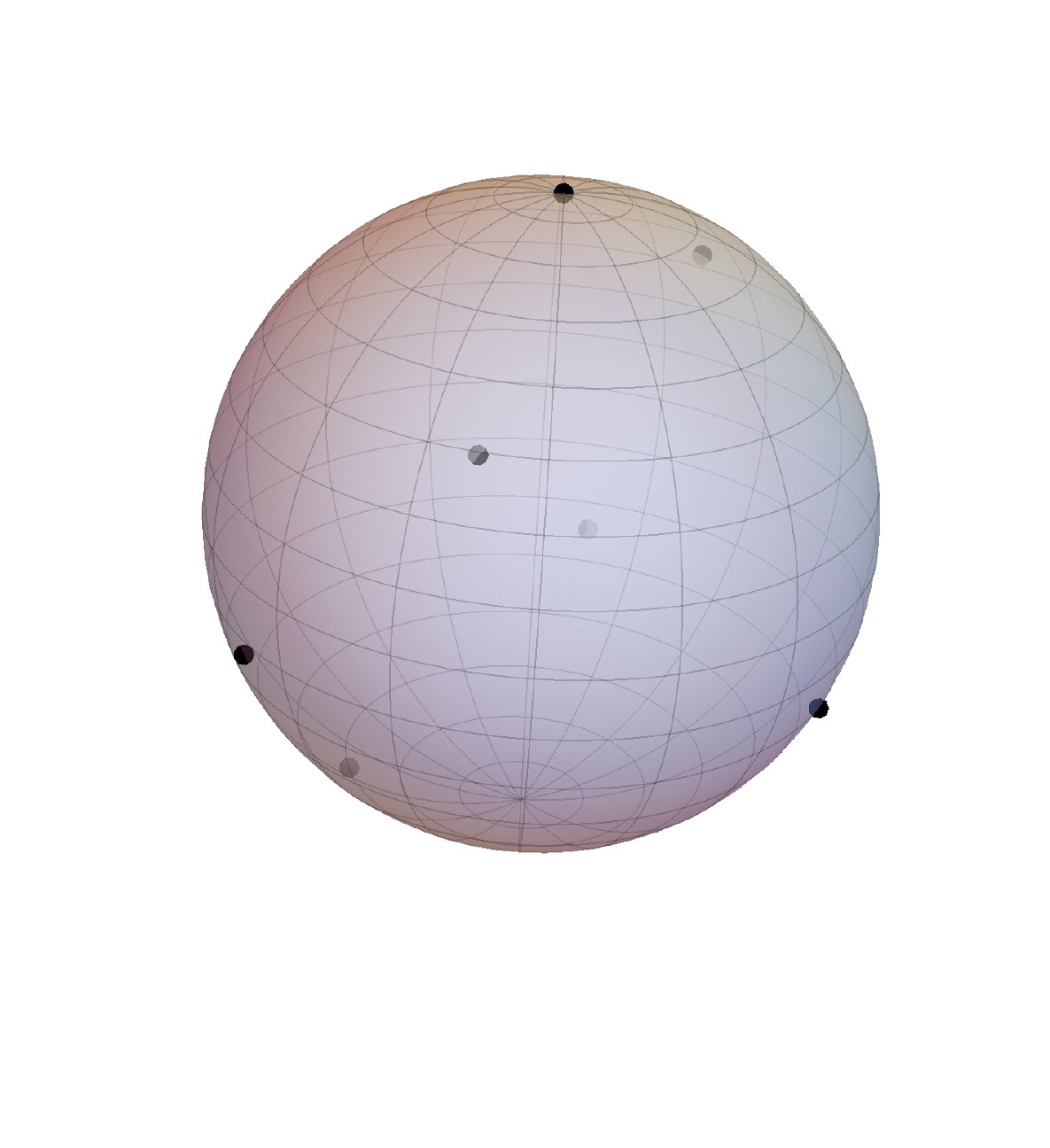}
    \qquad \vspace{-1cm}
    \includegraphics[scale=0.45]{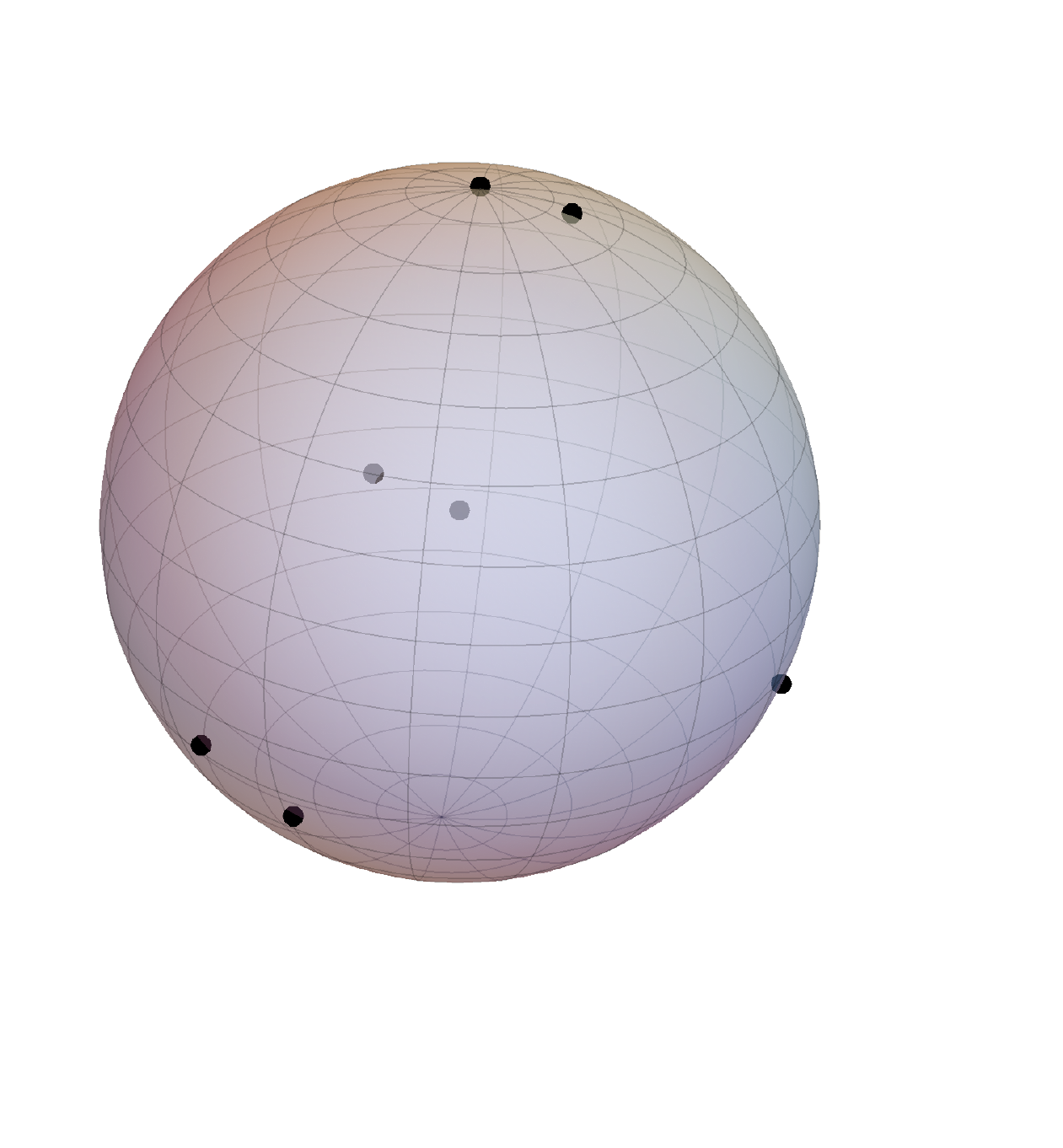}
    \caption{Left: maximal spin-$\frac{7}{2}$ state with four highlighted stars that approximate a regular tetrahedron.  Right: constrained optimization after snapping the four points to a regular tetrahedron. The right has a Wigner negativity of $0.73243$.}
    \label{fig:spin3p5ApproxTet}
\end{figure}

Using a method similar to the previous section, we have found another highly negative spin-7/2 state by modifying the maximal state.  Note that combining the isolated point with either of the triangles results in a triangular pyramid as shown in the middle and right plots of Fig.~\ref{fig:7qubits_combined}.  These pyramids are in general irregular in the sense that they are built from one equilateral face and three identical isosceles faces.  In our particular instance of this general pattern, the smaller pyramid has a characteristic isosceles face with angles $\{ 52^\circ, 52^\circ, 77^\circ \}$ and the larger has $\{ 62^\circ, 62^\circ, 56^\circ \}$.  The latter configuration is somewhat close to an equilateral triangle, which if true would turn the pyramid into a tetrahedron.  Pursuing this, we can consider the constrained problem where four of the seven stars are snapped to a tetrahedron while the remaining three are varied through the numerical optimization. The result of this is shown on the right of Fig.~\ref{fig:spin3p5ApproxTet}.  The constellation is similar to the maximal one, though different enough to be visually distinguished.  This state also has a high Wigner negativity of $0.73243$.  This differs from the true maximum slightly, within one part in a thousand.  We mention this because although this state is not maximal, it is nonetheless interesting to see a Platonic solid within an extremely negative state with spin $7/2$ despite the six-qubit extremal state not being the octahedron.

%width=0.45\textwidth

\section{Statistical rarity}

\begin{figure}
    \centering
    \includegraphics[scale=0.7]{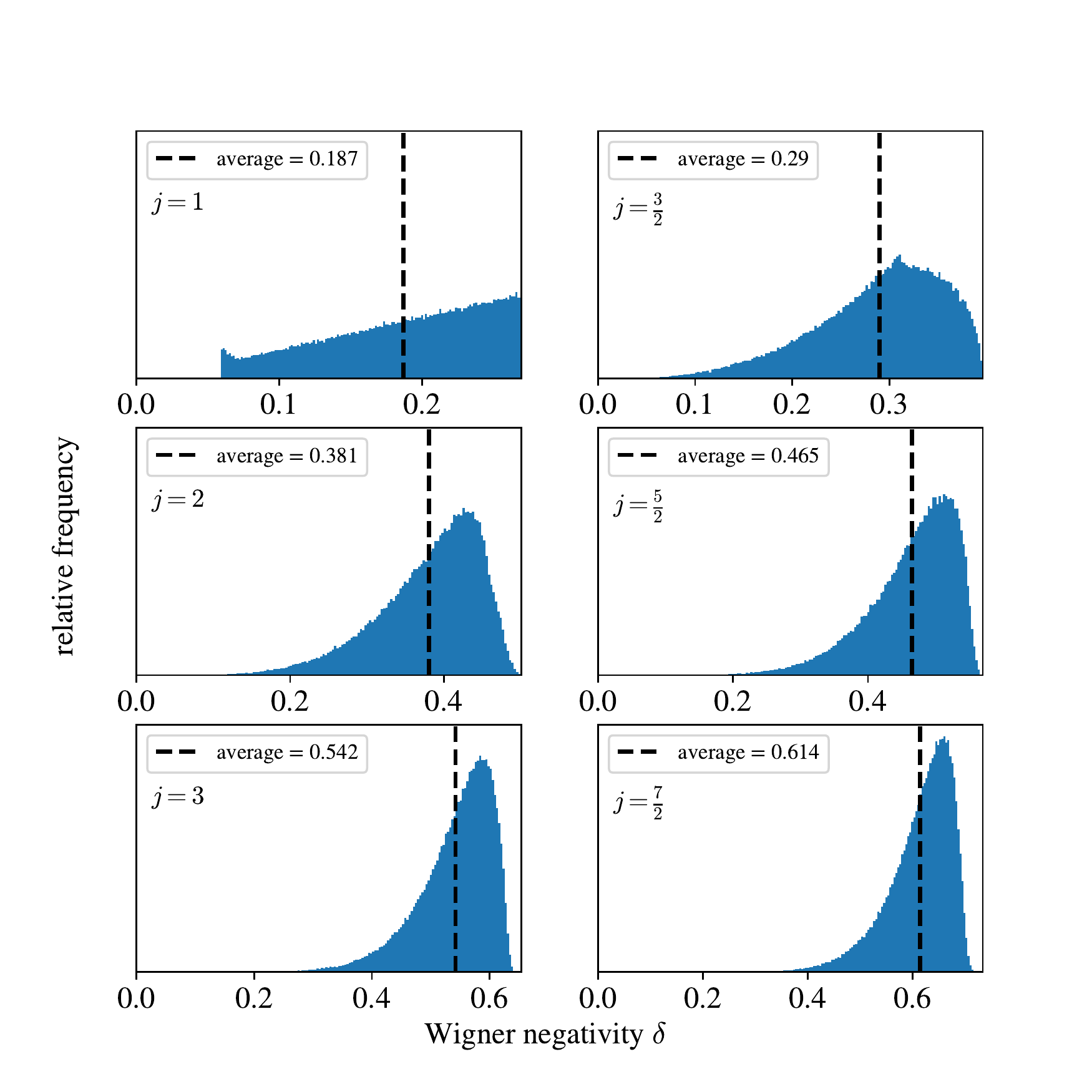}
    \caption{Distribution of Wigner negativity over $N=200000$ randomly selected states for spins $1 \le j \le 7/2$. The horizontal axis gives the absolute Wigner negativity, and cuts off on the right at the maximal value. In each case the vertical dashed line indicates the CUE average.}
    \label{fig:neg_hists}
\end{figure}

Will a randomly selected spin-$j$ state be likely to have a large amount of Wigner negativity, or a small amount?  This is the question we seek to address in this section.  Given a fixed spin $j$, random unitaries are sampled with respect to the Haar measure on U$(2j+1)$ -- i.e.\ the circular unitary ensemble (CUE).  These unitaries are applied to a fiducial state to produce a set of uniformly random states in the Hilbert space $\mathbb{C}^{2j+1}$, each having an associated constellation.  Fig.~\ref{fig:neg_hists} shows the distribution of Wigner negativity over $N = 200000$ random states for $1 \le j \le \frac{7}{2}$.  The results indicate that random states on average are highly Wigner negative relative to the allowed range, though the exact distribution depends on the particular spin.  Apart from the two and three qubit systems, each distribution has a similar form with a high peak relatively close to the upper bound.  As spin increases the peak narrows, creating a more rapid decay towards maximal negativity.

%width=0.95\textwidth
\begin{figure}
    \centering
    \includegraphics[scale=0.6]{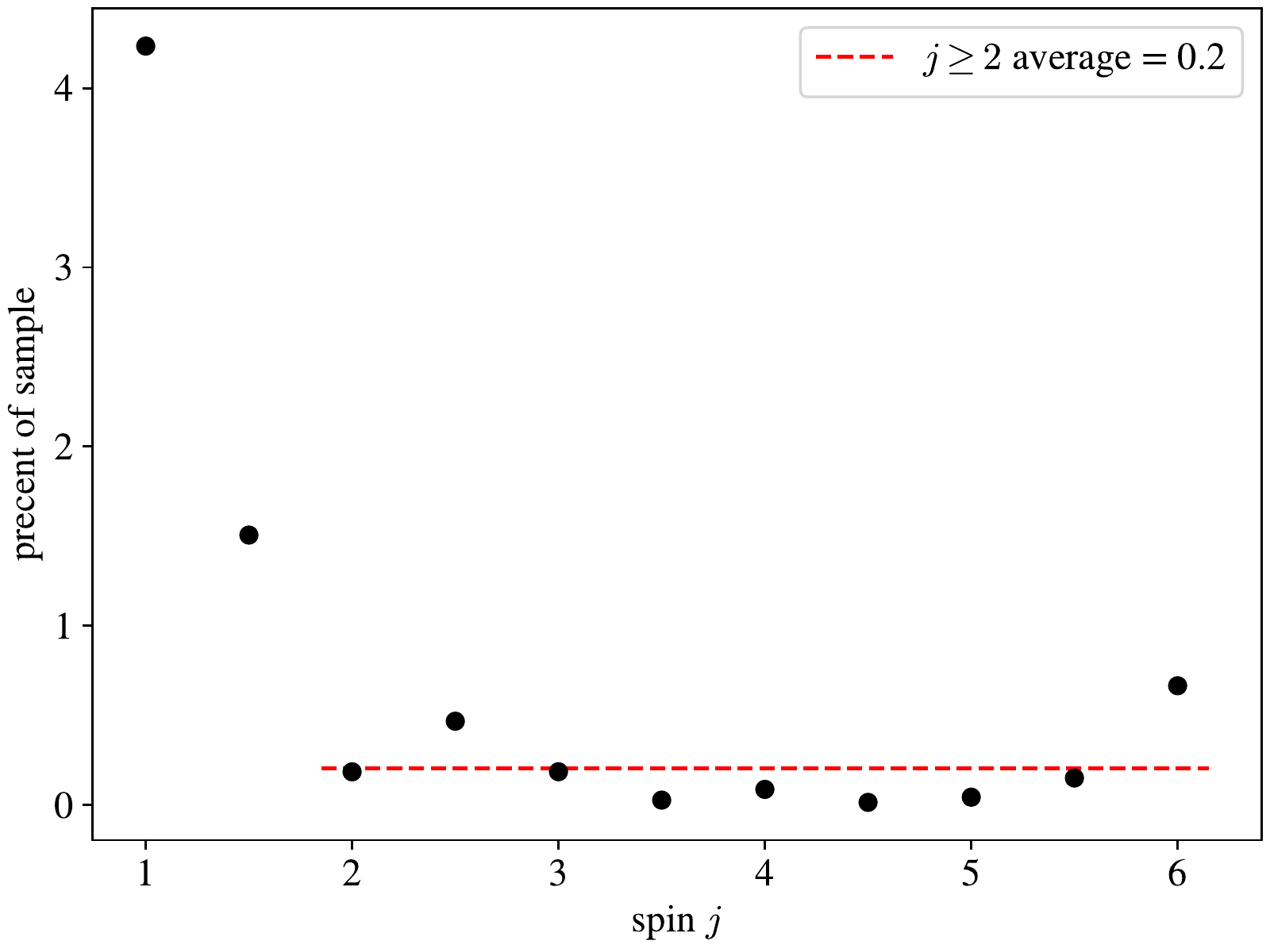}
    \caption{For each spin $j$ on the horizontal axis, the vertical axis is the percentage of the $N=200000$ random states that have a negativity within 2\% of their respective maximum.  The dashed red line is the average for $j \geq 2$.  For spins $j\geq 4$ the maximum is the highest negativity sampled.
    }
    \label{fig:within-2-percent}
\end{figure}

Additional distributions were computed for higher spins up to and including $j=6$, and they continue to have the same general form as the $j\geq 2$ set in Fig.~\ref{fig:neg_hists}.  The increasingly sharper decays from each peak indicate that the extremal state(s) and those with a similar degree of negativity become increasingly rare as spin increases.  Fig.\ \ref{fig:within-2-percent} plots the percentage of random states from the CUE sample that have a negativity within two percent of their theoretical maximum; see Appendix \ref{sec:numerical_data} for the list of maxima.  On average, across all spins $j \geq 2$, only $\approx 0.2$\% of states are within 2\% of their respective maximum negativity value, indicating the rarity of the highly nonclassical states in Hilbert space.  For spins $j\geq 4$ the maximum is taken to be the highest negativity sampled; the true extremal value can only be higher, making the average (dashed red line) almost certainly an upper bound.  By comparison, we computed the linear entropy of the ($1:n-1$) bipartition in the qubit representation of the same CUE sample and find that, on average across all spins $j \geq 2$, approximately $7.3$\% of them are within 2\% of the maximum value $1/2$.  This number increases with spin, with the case of $j=6$ having around 12.6\% of random states within 2\% of the linear entropy maximum.  In the specific case of an 11-qubit system for example, one is approximately 70 times more likely to randomly sample a state with an almost maximally mixed one-qubit reduced state than a state with an almost maximal Wigner negativity.

We also note that, as expected, all random states with $j < 4$ were found to have strictly less negativity than the determined maximal state of the same dimension.

\section{Discussion}

Determining the exact Wigner negativity of a state in general requires the identification of the zero set of its associated Wigner function, followed by an integration over that set.  This rapidly becomes analytically intractable as spin increases.  Even in the azimuthally symmetric case of Dicke states for example, the form of the Wigner function \eqref{eq:dicke-wigner} requires exact knowledge of the zeroes of the Legendre polynomials.  Hence we are left with computational techniques and general heuristics.  Here we discuss some observations in the context of states that maximize other measures of nonclassicality.  In particular, the constellations of such alternative maximal states are in general highly symmetric, highly delocalized, or both.  And while the Wigner-maximal constellations partially display these qualities in the spins considered, they do not follow an obvious geometric guiding principle.  

\begin{figure}
    \centering
    \includegraphics[scale=0.6]{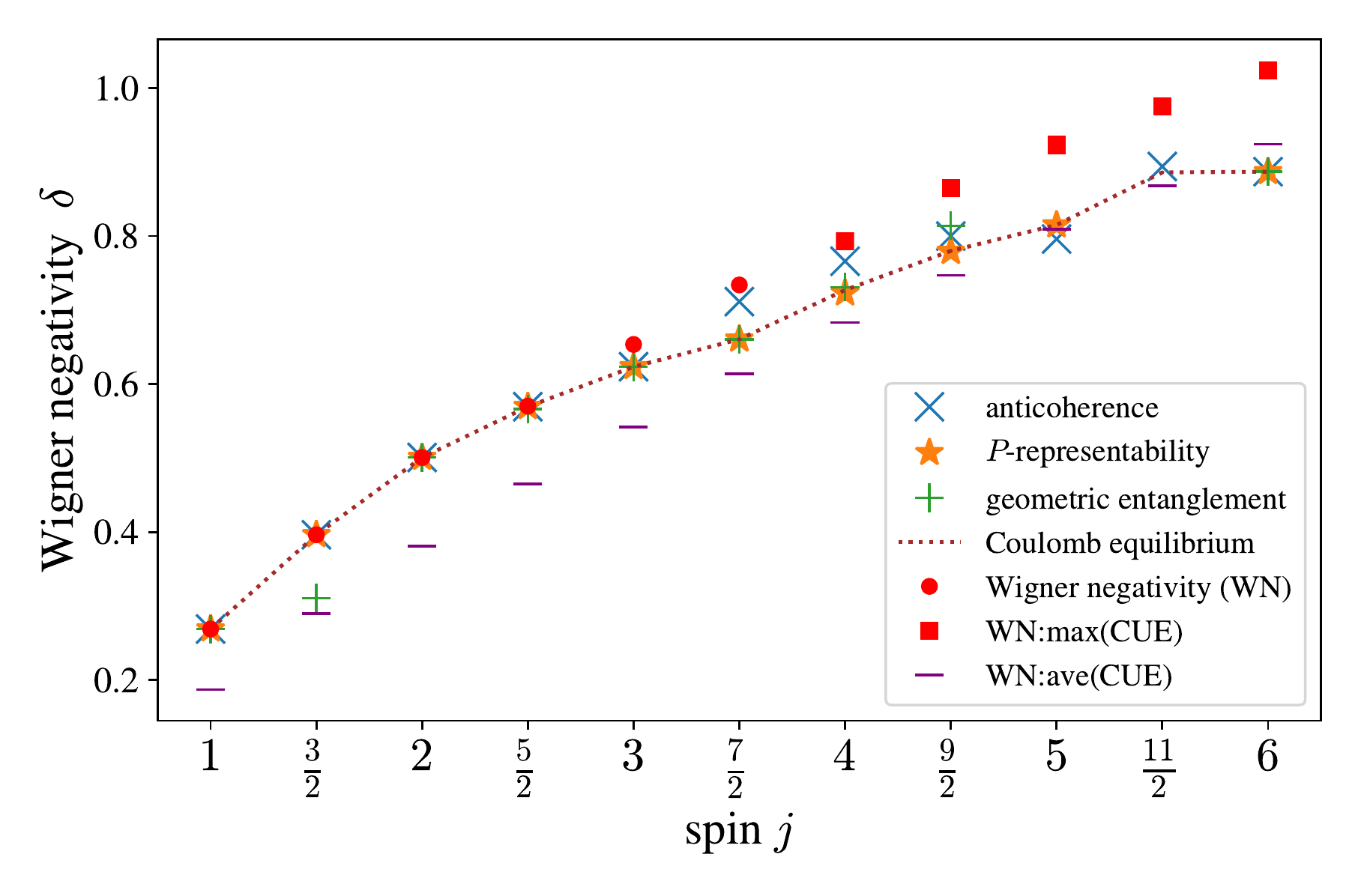}
    \caption{Wigner negativity of alternative maximal states and Thomson/Coulomb global equilibria.  For $j \geq 2.5$ (five qubits and larger), there exists a state with higher negativity than the other maxima considered.  For $j\geq 4$ this state is taken to be the most negative of the CUE random sample.  The lack of a marker indicates no available data.}
    \label{fig:alt-extremal-comparison}
\end{figure}

\begin{figure}
    \centering
    \includegraphics[width=0.45\textwidth]{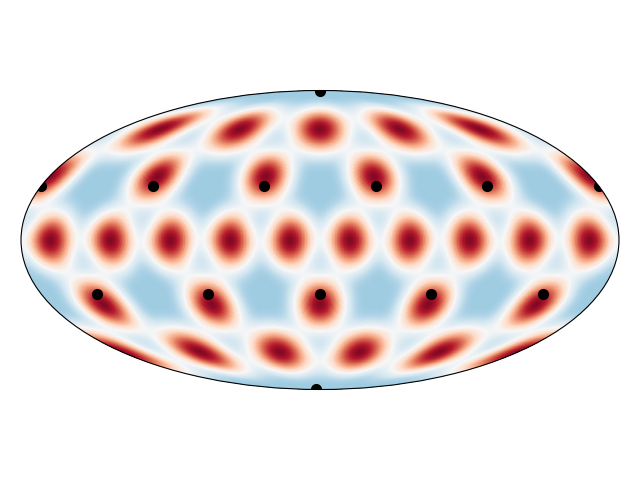}
    \quad
    \raisebox{0.0\height}{\includegraphics[width=0.45\textwidth]{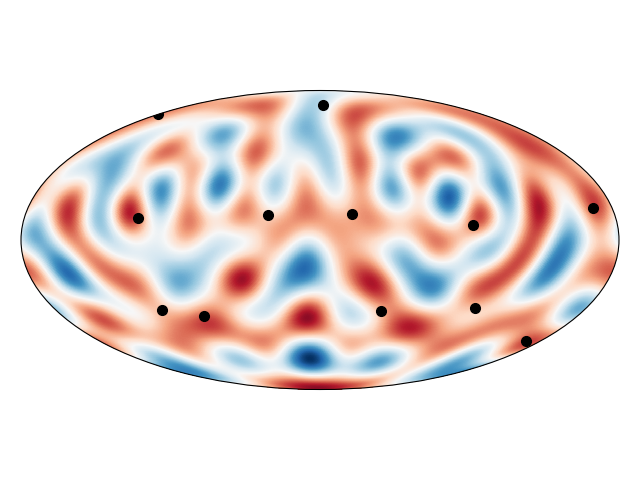}}
    \caption{Left: Wigner function of the icosahedron state.  Right: Wigner function of the most negative state found in the spin 6 CUE sample.}
    \label{fig:12-qubit-wigner-comparison}
\end{figure}

First consider constellation symmetry.  The relatively weak correlation between configuration symmetry and Wigner negativity begins with the spin 3 system where, despite the Wigner-maximal state having partial symmetry (i.e.\ a $C_{2\text{v}}$ point group), it is definitively not the highly symmetric octahedron state.  This continues to higher spin despite the global maximum being unknown.  See Fig.~\ref{fig:alt-extremal-comparison} for a comparison between all extremal states up to spin 6.  For $j \geq 4$ the most negative random state drawn from the CUE sample is used in place of the unknown global maximum(s).  Only for $j \leq 2$ does the highest Wigner-negative state coincide with one of the alternative maxima.  Each larger dimensional system with $j>2$ contains at least one state in Hilbert space with a Wigner negativity larger than the negativity of the alternative extremal states.  This difference in negativity furthermore appears to grow with spin.  The last case of spin 6, corresponding to 12 indistinguishable qubits, is particularly interesting because that is the number of vertices of the icosahedron, another Platonic solid with high constellation symmetry.  Indeed the ``icosahedron state" simultaneously maximizes the other measures of nonclassicality yet is approximately $\approx13\%$ less negative than the statistical maximum, and is actually below average across the Hilbert space $\mathbb{C}^{13}$.  Fig.~\ref{fig:12-qubit-wigner-comparison} compares the icosahedron state and the most negative random state sampled.  There is a passing similarity between the Wigner functions of the 12 qubit statistical maximum and both the 6 and 7 qubit global maxima, with two roughly dual ``lobes" in the upper hemisphere together with somewhat concentric regions in the opposing hemisphere.  It is plausible that the 12-qubit global maximum may sharpen this similarity and demonstrate a more concrete pattern in Wigner-maximal spin states.

We also briefly mention the two other Platonic states within $1/2 \leq j \leq 6$: the cube and the tetrahedron.  The cube state is interesting because anticoherence is the only measure that witnesses it as extremal.  It is additionally the first time all four measures have different maximal states.  In contrast, the tetrahedron state is the only non-trivial case where all four measures agree.  This consensus offers evidence that the tetrahedron state may be of practical use in various quantum-enhanced applications.  This is also in part what motivated us to pursue modifications of the maximal states revolving around fixed tetrahedra within a given constellation.

Now consider constellation delocalization.  To compare the Wigner-maximal states against those with a uniform distribution of Majorana stars we consider the Thompson problem, defined as the electrostatic configuration of $n$ point charges confined to the sphere that minimizes the Coulomb potential energy \cite{Whyte_1952}.  The solutions to the Thompson problem are one of many inequivalent benchmarks for distributional uniformity over the sphere \cite{saff1997distributing}, however the intuitive description of the problem makes it a reliable reference point.  The exact configurations to this problem are not generally known for an arbitrary number of points, though numerically optimized solutions exist in many cases  \cite{sloane_thomson, Wales_Ulker}.  The dotted line in Fig.~\ref{fig:12-qubit-wigner-comparison} denotes the Wigner negativity of such Thomson solutions.  Similar to the case of constellation symmetry, the Coulomb equilibria have sub-maximal negativity for $j > 2$.  The difference between the equilibria and the highest known negativity also grows with system dimension, culminating again in the below-average negativity from the spin 6 icosahedron state.

\begin{figure}
    \centering
    \includegraphics[scale=0.6]{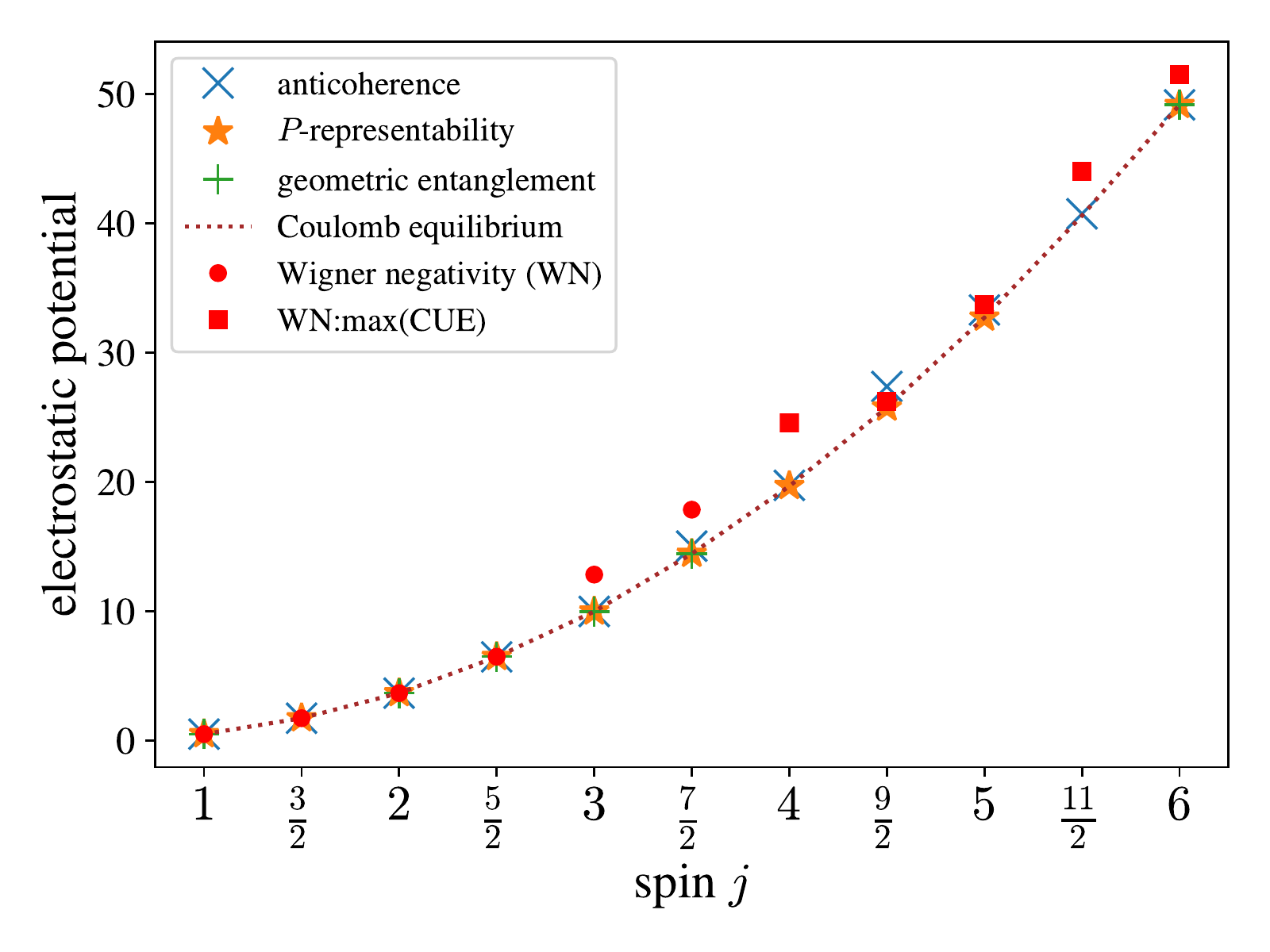}
    \caption{Coulomb electrostatic potential energy of alternative extremal constellations.  Highly Wigner-negative states tend to have a higher potential energy, indicating the presence of clustering of stars within their constellations.  The lack of a marker indicates no available data or, as in some cases of geometric entanglement, when two stars are degenerate.}
    \label{fig:coulomb-potential}
\end{figure}

We note a related observation that Wigner-maximal constellations sometimes contain groups of stars confined to a relatively small region of phase space.  This is seen in the small edges of the rectangle structure within spin 3 (Fig.~\ref{fig:spin3Max}), and the two triangles within spin 7/2 (Fig.~\ref{fig:spin3p5Max}).  Qualitatively, such grouping is not generally seen in the non-Wigner maximal states, with the exception of geometric entanglement in the spin-3/2 case having two stars degenerate on the same point.  The geometric measure configurations for spins greater than $7/2$ continue to occasionally have degenerate stars \cite{Aulbach_2010}, however they are still relatively uniform over the sphere if such degenerate points are seen as singular.  This ``clustering" of stars, i.e.\ being close but not degenerate, appears to be specific to Wigner negativity.  See Fig.~\ref{fig:coulomb-potential} for a comparison of all the maximal states as measured by the electrostatic potential energy of their constellations.  States with high negativity tend to have higher potential energy, indicating the presence of relatively closer stars.

In summary, given the data from the spins considered, it appears that while Wigner negativity is sensitive to both constellation symmetry and delocalization, they are not guiding principles to be individually optimized over.

%3qubit ge = 0.310
%3qubit K,Q,WN = 0.396
%3qubit average = 0.29

%4qubit ge,K,Q,WN = 0.501
%4qubit average = 0.381

%5qubit thomson, K, Q = 0.5694
%5qubit GE = 0.5662
%5qubit WN = 0.5702
%5qubit average = 0.465

%6qubit thomson, ge, Q, K = 0.6233
%6qubit WN = 0.6535
%6qubit average = 0.542

%7qubit thomson, GE, Q = 0.660
%7qubit K = 0.711
%7qubit WN = 0.734
%7qubit average = 0.614

% ----- stats only -------
%8qubit average = 0.683
%8qubit stat max = 0.79332
%8qubit Kings = 0.766
%8qubit Queens = 0.723
%8qubit geom = 0.731

%9qubit average = 0.747
%9qubit stat max = 0.8648
%9qubit Kings = 0.8
%9qubit Queen = 0.77891
%9qubit geom = 0.814

%10qubit average = 0.809
%10qubit stat max = 0.9225
%10qubit King = 0.796

%11qubit average = 0.868
%11qubit stat max = 0.9747
%11 qubit king = 0.894

%12qubit average = 0.924
%12qubit stat max = 1.024
%icosahedron = 0.887

\section{Conclusion}

We have identified the maximally Wigner-negative quantum states in dimensions $\{2,\,\cdots,8\}$ as measured by the SU(2)-covariant Stratonovich-Wigner function on a spherical phase space.  These states have been characterized by their stellar representation and compared to three alternative notions of nonclassicality also studied in the context of spin: anticoherence, geometric entanglement, and $P$-representability.  For systems made of 5, 6, and 7 indistinguishable qubits there is complete disagreement with the other measures.  This is especially noteworthy for the case of 6 qubits where the octahedron state, despite simultaneously maximizing the other measures, is not the most Wigner-negative configuration.  The maximal constellations were also seen to display a local clustering of stars within a relatively small region of phase space, something that appears to be unique to the Wigner approach.

Higher dimensions were analyzed statistically, based on random CUE sampling over Hilbert space.  The results suggest that the departure from the other extremal states appears to grow with spin.  Random state statistics also showed that while the average negativity of states across Hilbert space is relatively high, states with utlra-high negativity are extremely rare.  This is in contrast with entanglement entropy across the constituent qubits.  The case of 12 qubits is notable because the most negative random state sampled is significantly more negative than the icosahedron state, which in fact has a negativity value below average across its Hilbert space despite simultaneously saturating the alternative measures of nonclassicality.  These results suggest that neither a purely symmetry-based or purely delocalization-based geometric principle in the constellations may fully capture the nonclassicality of spin states as measured by Wigner negativity. Furthermore, our findings identify interesting new nonclassical states that could be worth exploring as resources for quantum information processing. 

An important additional result is a proof that all spin coherent states of arbitrary dimension must take negative values in their Wigner function.  This, together with numerical evidence that such states also minimize negativity, offers strong evidence that all finite dimensional pure spin states have non-vanishing SU(2)-covariant Wigner negativity.  This likely non-existence of Gaussian/stabilizer analogues would be unique to spin as compared against a Heisenberg-like dynamical symmetry, and is of interest in the context of multiqubit systems and their optimal use for quantum information processing applications.

\section*{Acknowledgements}
This work was supported by the Natural Sciences and Engineering Research Council of Canada (NSERC). The work of R.~A.~H.~was supported in part of the Natural Sciences and Engineering Research Council of Canada (NSERC), Asian Office of Aerospace Research and Development Grant No. FA2386-19-1-4077, and received the support of a fellowship from ``la Caixa” Foundation (ID 100010434) and from the European Union’s Horizon 2020 research and innovation
programme under the Marie Marie Skłodowska-Curie grant agreement No 847648 under fellowship code LCF/BQ/PI21/11830027. J.D. acknowledges discussions with Aaron Goldberg and Meenu Kumari. We thank Keith Ng for computational assistance. 
Wilfrid Laurier University and the University of Waterloo are located in the traditional territory of the Neutral, Anishnawbe and Haudenosaunee peoples. We thank them for allowing us to conduct this research on their land. 

\appendix

\section{Alternative definitions of nonclassicality}\label{sec:alternative_descriptions}
There are many other definitions of nonclassicality, particularly in the SU(2) setting where the respective maximally nonclassical state(s) may be visualized and compared through the stellar representation.  Here we briefly review a non-exhaustive list of such definitions used in spin systems.

\textit{Anticoherence} is a natural approach to quantify how different a state is to a spin coherent state.  An anticoherent state of order $M$ is generally defined as a spin state having zero average spin, $\langle \bm{J} \rangle = 0$, as well as isotropic higher moments, $\langle (\bm{n}\cdot \bm{J})^k \rangle \neq f(\bm{n}) \,\forall \, k=1, ..., M$ \cite{zimba2006anticoherent}.  A more refined definition follows from the multipole expansion of a density matrix $\rho$ with fixed spin $j$,
\begin{equation}
    \rho = \sum_{K=0}^{2j} \sum_{q=-K}^K \rho_{Kq} T^{(j)}_{Kq}, \qquad \rho_{Kq} = \tr[\rho T^{(j) \dagger}_{Kq}],
\end{equation}
where $T^{(j)}_{Kq}$ are the spherical tensor operators \eqref{eq:spherical_tensor}.  The state multipoles $\rho_{Kq}$ contain information on the amplitude of a density matrix to have a specific multipole pattern, and the quantity $\sum_q |\rho_{Kq}|^2$ is the overlap with the entire $K$ multipole.  Higher $K$ reflects finer angular structure, so it is natural to analyze the cumulative overlap
\begin{equation}\label{eq:anticoherence}
    \mathcal{A}^{(j)}_M = \sum_{K=1}^M \sum_{q=-K}^k |\rho_{Kq}|^2.
\end{equation}
Spin-$j$ coherent states are known to maximize the above quantity for all orders $M$ (i.e. they have the strongest collective polarization allowed), and so states that minimize Eq.\ \eqref{eq:anticoherence} are interpreted as spin coherent ``opposites".  States for which $ \mathcal{A}^{(j)}_M $ in Eq.\ \eqref{eq:anticoherence} vanish are said to be $M$th-order unpolarized, with their polarization information having been pushed to higher multipoles.  In general such states are called the Kings of Quantum.  Various King states of spin $j$ and order $M$ have been calculated \cite{bjork_kings_2015}, experimentally realized \cite{Bouchard_2017}, and are of critical metrological use in achieving the Heisenberg bound in quantum rotosensing where both the rotation angle and rotation axis are unknown \cite{goldberg_james_euler_2018,Goldberg_rotosensing_limits_2021}.  In the qubit picture, a spin state $\ket{\Psi_S}$ being $M$-anticoherent is equivalent to the reduced state of $h$ qubits $\rho_h = \tr_{2j-h}(\ketbra{\Psi_S}{\Psi_S})$ being maximally mixed for all $h = 1,...,M$ \cite{mixed-1qubit-reductions_bastin_2014}.

In an optical setting, anticoherence is about quantifying the quantum mechanical departure from the classical fact that fully polarized light always has a Stokes vector $(S_x, S_y, S_z)$ on the Poincare sphere.  Higher intensity beams continuously enlarge the Poincare radius but not the angular information of the Stokes vector.  In the quantized picture the Poincare radii become discrete, and higher layers permit the vanishing of polarization expectation $(\langle J_x \rangle, \langle J_y \rangle, \langle J_z \rangle)$ even for pure states \cite{Goldberg_polarization_review_2021}.  Intuitively, each photon in a multi-photon pure state of light may have its own polarization (Majorana star), allowing for the possibility of collective cancellation.

The \textit{geometric measure of entanglement}, $E_G$, is a widely used entanglement monotone introduced by Shimony as the smallest distance to the set of product states \cite{Shimony_1995}.  Here we express the definition in the context of SU(2) symmetry where the set of product states becomes the set of spin coherent states \cite{Hubener_symmetric_product_geometric_2009}:
\begin{equation}\label{eq:geo_definition_Shimony}
    E_G (\ket{\psi}) = \frac{1}{2} \min_{\ket{\phi} \in \mathcal{C}_{\text{scs}}} || \ket{\psi} - \ket{\phi} ||^2_{\text{HS}},
\end{equation}
where $||A||_{\text{HS}} = \sqrt{\tr[A^\dagger A]}$ is the Hilbert-Schmidt norm and
\begin{align}
    \mathcal{C}_{\text{scs}} &= \{ \ket{\chi}^{\otimes 2j}, \,\, \ket{\chi} = \cos{\frac{\theta_\chi}{2}}\ket{0} + e^{i\phi_\chi}\sin\frac{\theta_\chi}{2}\ket{1} \} \nonumber \\
    &= \{ \ket{\theta,\phi},\, \, (\theta,\phi) \in S^2 \}
\end{align}
is the set of product (i.e.\ ``classical") states.  Combining Eq.\ \eqref{eq:geo_definition_Shimony} with the fact that the $(s=-1)$ SU(2) kernel is the spin coherent POVM over the sphere,
\begin{equation}\label{eq:husimi-povm}
    Q_\rho (\theta,\phi) := f^{(-1)}_\rho (\theta,\phi) = \tr[\, \rho \, \ketbra{\theta,\phi}{\theta,\phi} \, ] = \langle \theta,\phi | \rho | \theta,\phi \rangle,
\end{equation}
together shows that 
\begin{equation}\label{eq:alternate-geo-using-husimi}
    E_G(\ket{\psi}) = \min_{(\theta,\phi)\in S^2} (1 - Q_\psi(\theta,\phi)) = 1 - \max_{(\theta,\phi)\in S^2} Q_\psi(\theta,\phi).
\end{equation}
Hence the geometric measure of a state may be thought of as the relative difference between the maximum height of the state's Husimi function and its theoretical upper bound of unity.  Consequently, maximally geometric-entangled pure states may be thought of as those with the ``flattest" possible Husimi function \cite{Dodonov_2000}.

$P$\textit{-representability} is similar to the geometric measure of entanglement, but with the set of states deemed classical enlarged to the convex hull of spin coherent states \cite{Giraud_2008_classicality,Giraud_2010}.  With the Glauber-Sudarshan $P$ function seen as the collection of expansion coefficients over the spin coherent projector basis, the classical set is comprised of those states that admit a positive $P$-function:
\begin{equation}
    \mathcal{P}(\rho) = \min_{\rho_c \in \mathcal{C}} || \rho - \rho_c ||_{\text{HS}}
\end{equation}
where, using the appropriate Stratonovich kernel, Eq. \eqref{eq:SU(2)-kernel-diagonal},
\begin{equation}
    \mathcal{C} = \{ \rho \, | \, P_\rho (\theta,\phi) = \tr[\rho \Delta_{SU(2)}^{(1)} (\theta,\phi) ] \geq 0 \quad \forall \, (\theta,\phi) \in S^2 \}
\end{equation}
for a given spin $j$.  An alternative way to compare the geometric measure and $P$-representability is to keep the convex hull fixed; then the geometric measure is effectively seen as minimizing the Bures distance rather than the Hilbert-Schmidt distance \cite{martin_giraud_geometric_high_2010}.  In any case, the motivation behind this notion of nonclassicality comes from interpreting the values of a positive $P$-function as a collection of epistemic/statistical weights in an incoherent mixture of spin coherent states.  States that maximize this measure are known as the Queens of Quantum \cite{Giraud_2008_classicality, Giraud_2010}.

%It may be worth mentioning that the above three notions in some sense \textit{a priori} consider a set of states deemed as ``classical", with nonclassicality following as a comparison against that set.  Wigner negativity on the other hand is \textit{a priori} agnostic to which states may or may not be nonclassical.  That spin coherent states appear to be the least nonclassical according to Wigner negativity is a conclusion and not an assumption, implicit or explicit.

\section{Numerical data on maximal states}\label{sec:numerical_data}

% \begin{table*}
% \begin{ruledtabular}
% \begin{tabularx}{\textwidth}{cc}
% %\begin{tabularx}{\textwidth}{ccp{0.1\linewidth}p{0.1\linewidth}c}
% Spin &  Dicke Representation 
% \\ \hline
% $1$ & $\ket{1,0}$
% \\  \hline
% $3/2$ &  $\tfrac{1}{2} \ket{\tfrac{3}{2}, \tfrac{3}{2}} - \sqrt{\tfrac{3}{2}} \ket{\tfrac{3}{2}, -\tfrac{1}{2}}$
% \\ \hline
% $2$ & $\sqrt{\tfrac{1}{3}} \ket{2, 2} + \sqrt{\tfrac{2}{3}} \ket{2, 1}$
% \\ \hline 
% $5/2$ & $0.216 \ket{\tfrac{5}{2}, \tfrac{5}{2}} - (0.391 - 0.507i) \ket{\tfrac{5}{2}, \tfrac{3}{2}} + (0.053 + 0.200i) \ket{\tfrac{5}{2}, \tfrac{1}{2}} + (0.090 + 0.034 i) \ket{\tfrac{5}{2}, -\tfrac{1}{2}} - (0.594-0.373i) \ket{\tfrac{5}{2}, -\tfrac{3}{2}}$
% \end{tabularx}
% \end{ruledtabular}
% \caption{Table of Dicke coefficients Wigner extremal states.}
% \label{master-table}
% \end{table*}

\begin{table*}[h]
\begin{ruledtabular}
\begin{tabularx}{\textwidth}{ccc}
%\begin{tabularx}{\textwidth}{ccp{0.1\linewidth}p{0.1\linewidth}c}
Spin &  Wigner Negativity & Extremal constellation $(\theta, \phi)$
\\ \hline
$1$ & 0.26935 & $(0,0), (\pi,0)$
\\  \hline
$3/2$ & 0.39634 & $(0,0), (2\pi/3,0), (2\pi/3,\pi)$
\\  \hline
$2$ & 0.50078 & $(0,0), (\theta_T, 0), (\theta_T, 2\pi/3), (\theta_T, 4\pi/3)$\footnote{$\theta_T = 2 \cos^{-1}(1/\sqrt{3})$} 
\\ \hline
$5/2$ & 0.57016 & $(0,0),(1.66,0),(1.43,2.21),(2.86,2.23),(1.65,4.43)$
\\ \hline
$3$ & 0.65354 & $(0,0), (1.62,0), (1.71,2.03), (1.71,4.25), (2.02,4.54), (2.02,1.75)$
\\ \hline
$7/2$ & 0.73395 & $(0,0),(1.97,0),(1.83,2.18),(2.07,4.51),(1.83,4.09),(2.06,1.76),(0.43,6.25)$
\end{tabularx}
\end{ruledtabular}
\caption{Table of maximal Wigner negative values and the associated constellations.}
\label{master-table}
\end{table*}

\begin{table*}[h]
\begin{ruledtabular}
\begin{tabularx}{\textwidth}{cc}
%\begin{tabularx}{\textwidth}{ccp{0.1\linewidth}p{0.1\linewidth}c}
Spin &  Dicke coefficients $(m=-j, \cdots, j)$
\\ \hline
$1$ & $(0,1, 0)$
\\  \hline
$3/2$ &  $(0, -\sqrt{3/2}, 0, 1/2)$
\\ \hline
$2$ & $(0, \sqrt{2/3},0,0,1/\sqrt{3})$
\\ \hline 
$5/2$ & $(0, -0.594 + 0.373 i, 0.090 + 0.034 i, 0.053 + 0.200i, -0.391 + 
 0.507i, 0.216)$
\\ \hline
$3$ & $(0, 0.743 - 0.001 i, -0.02, 0.156, 0.37, -0.111, 0.523)$
\\ \hline 
$3/2$ & $(0, 0.299 - 0.008 i, 0.687 - 0.006 i, -0.227 - 0.005 i, 0.299 - 
 0.001 i, 0.215 - 0.003 i, -0.074 - 0.005 i, 0.496)$
\end{tabularx}
\end{ruledtabular}
\caption{Table of Dicke coefficients Wigner extremal states.  Exact numbers are used when available.}
\label{master-table}
\end{table*}

\newpage
\bibliography{extremal-states}

\end{document}